\documentclass[12pt]{article}
\pdfoutput=1
\usepackage{jheppub}
\usepackage{subcaption}
\usepackage{amsmath}
\usepackage{amssymb}
\usepackage{pifont}

\usepackage{amsfonts}

\def\d{\operatorname{d}}

\newcommand{\nn}{\nonumber}

\def\bal#1\eal{\begin{align}#1\end{align}}

\def\secnum[#1]{\texorpdfstring{$#1$}{TEXT}}

\def\secnuml#1\secnumr{\texorpdfstring{$#1$}{TEXT}}

\def\im{\text{i}}
\def\eqa{\begin{eqnarray}}
\def\eqae{\end{eqnarray}}
\def\eq{\begin{equation}}
\def\eqe{\end{equation}}
\def\be{\begin{equation}}
\def\ee{\end{equation}}
\def\bea{\begin{eqnarray}}
\def\eea{\end{eqnarray}}
\def\ba{\begin{array}}
\def\ea{\end{array}}
\def\bd{\begin{displaymath}}
\def\ed{\end{displaymath}}

\def\Tr{{\rm Tr}}
\def\tr{{\rm tr}}
\def\>{\rangle}
\def\<{\langle}

\def\k{\kappa}

\def\G{\Gamma}

\def\L{\Lambda}

\def\S{\Sigma}

\def\({\left(}
\def\){\right)}
\def\nn{\nonumber \\}

\title{More on Half-Wormholes and Ensemble Average}

\author{Jia Tian and Yingyu Yang}
\affiliation{Kavli Institute for Theoretical Sciences (KITS),\\
University of Chinese Academy of Sciences (UCAS), Beijing 100190, China}
\emailAdd{wukongjiaozi@ucas.ac.cn, yangyingyu18@mails.ucas.ac.cn  }

\abstract{We continue our study \cite{Peng:2022pfa} about the half-wormhole proposal. By generalizing the original proposal of half-wormhole we propose a new way to detect half-wormholes. The crucial idea is to decompose the observables into self-averaged sector and non-self-averaged sectors. We find the contributions from different sectors have interesting statistics in the semi-classical limit. In particular, dominant sectors tend to condense and the condensation explains the emergence of half-wormholes and we expect that the appearance of condensation is a signal of possible bulk description. 
We also initiate the study of multi-linked-half-wormholes using our approach. }

\begin{document}

\today

\maketitle

\section{Introduction}
Recent progress in quantum gravity and black hole physics impresses on the fact that wormholes play important roles\footnote{For a up to date review, see \cite{Kundu:2021nwp}}. Many evidences suggest an appealing conjectural duality between a bulk gravitation theory and an ensemble theory on the boundary \cite{Saad:2019lba,Stanford:2019vob, Iliesiu:2019lfc, Kapec:2019ecr,Maxfield:2020ale, Witten:2020wvy, Arefeva:2019buu, Betzios:2020nry, Anninos:2020ccj, Berkooz:2020uly, Mertens:2020hbs, Turiaci:2020fjj, Anninos:2020geh, Gao:2021uro, Godet:2021cdl, Johnson:2021owr, Blommaert:2021etf, Okuyama:2019xbv,Forste:2021roo, Maloney:2020nni, Afkhami-Jeddi:2020ezh,Cotler:2020ugk,Benjamin:2021wzr,Perez:2020klz,Cotler:2020hgz,Ashwinkumar:2021kav,Afkhami-Jeddi:2021qkf,Collier:2021rsn, Benjamin:2021ygh,Dong:2021wot,Dymarsky:2020pzc,Meruliya:2021utr,Bousso:2020kmy,Janssen:2021stl,Cotler:2021cqa,Marolf:2020xie,Balasubramanian:2020jhl,Gardiner:2020vjp,Belin:2020hea,Belin:2020jxr,Altland:2021rqn,Belin:2021ibv,Peng:2021vhs,Banerjee:2022pmw,Johnson:2022wsr,Collier:2022emf,Chandra:2022bqq,Schlenker:2022dyo,Kruthoff:2022voq,Kar:2022vqy,Cotler:2022rud}. For example  the seminal work \cite{Saad:2019lba} shows that Jackiw-Teitelboim (JT) gravity is equivalent to a random matrix theory. On the other hand this new conjectural duality is not compatible with our general belief about the AdS/CFT correspondence. A sharp tension is the puzzle of factorization \cite{Witten:1999xp,Maldacena:2004rf}. In \cite{Saad:2021uzi}, this puzzle is  studied within a toy model introduced in~\cite{Marolf:2020xie}, where they find that (approximate) factorization can be restored if other saddles which are called half-wormholes are included. Motived by this idea, in~\cite{Saad:2021rcu} a half-wormhole saddle is proposed  in a 0-dimensional (0d) SYK model , followed by further analyses in different models~\cite{Mukhametzhanov:2021nea,Garcia-Garcia:2021squ, Choudhury:2021nal, Mukhametzhanov:2021hdi,Okuyama:2021eju,Goto:2021mbt,Blommaert:2021fob,Goto:2021wfs}. In our previous works \cite{Peng:2021vhs,Peng:2022pfa}, we pointed out the connection between the gravity computation in~\cite{Saad:2021uzi} and the field theory computation in~\cite{Saad:2021rcu} and tested the half-wormhole proposal in various models. The main difficulty of this proposal is the construction of the half-wormhole saddles. Further more the ansatz proposed in \cite{Saad:2021rcu,Mukhametzhanov:2021hdi} seems to rely on the fact the ensemble is Gaussian with zero mean value. As a result, the 0d SYK model only has non-trivial cylinder wormhole amplitude. However for a generic gravity theory for example the JT gravity, disk and all kinds of wormhole amplitudes should exist. In our previous work \cite{Peng:2022pfa}, we find even turning on disk amplitude in 0d SYK model will change the half-wormhole ansatz dramatically. 

In this work, we generalize the idea of \cite{Saad:2021uzi} and propose a method of searching for half-wormhole saddles. In our proposal, the connection between \cite{Saad:2021uzi} and \cite{Saad:2021rcu} will manifest. One notable benefit of our approach is that it does not depend on the trick of introducing a resolution identity used in \cite{Saad:2021rcu}, the collective variables emerge automatically. More importantly our proposal can be straightforwardly generalized to non-Gaussian ensemble theories.

\section{Gaussian distribution or the CGS model}
In \cite{Saad:2021uzi}, the main model is the Coleman and Giddings-Strominger (CGS) model.
The CGS model is a toy model of describing spacetime wormholes and it is more suggestive to obtain it from the Marolf-Maxfield (MM) model \cite{Marolf:2020xie} by restricting the sum over topologies to only include the disk and the cylinder \cite{Saad:2021uzi}. 

Let the amplitudes of the disk and cylinder be $\mu$ and $t^2$, i.e.
\bea 
\langle \hat{Z}\rangle=\mu,\quad \langle \hat{Z}^2\rangle-\langle \hat{Z}\rangle^2=t^2,
\eea 
where $|\rangle=|\text{HH}\rangle$ denotes the no-boundary (Hartle-Hawking) state and $\hat{Z}$ denotes the boundary creation operator thus $\langle \hat{Z}^n\rangle$ computes the Euclidean path integral over all manifolds with $n$ boundaries.  For CGS model the gravity amplitude or the ``correlation function of the partition function"  $\langle \hat{Z}^n\rangle$ is a polynomial of $\mu$ and $t^2$ and in particular its generating function is simply
\bea 
\langle e^{u \hat{Z}}\rangle=\exp\({ u \mu+\frac{u^2t^2}{2}}\).
\eea 
Thus we can identify $\hat{Z}$ as a Gaussian random variable $Z$ such that the gravity amplitude $\langle f(\hat{Z})\rangle$ can be computed as the ensemble average $\mathbb{E}(f(Z))\equiv \langle f(Z)\rangle$. This equivalence is a baby version of gravity/ensemble duality.\par 
The crucial idea of \cite{Saad:2021uzi} is that the correlation functions of partition function does not factorize in general but they factorize between $\alpha$--states which are the eigenstates of $\hat{Z}$
\bea \label{inz2}
\langle \alpha|\hat{Z}^2|\alpha\rangle=\langle \alpha|\hat{Z}|\alpha\rangle^2=Z_\alpha^2.
\eea 
The $\alpha$-state is also created by a generation operator acting on $|\text{HH}\rangle$
\bea 
|\alpha\rangle=\psi_\alpha|\text{HH}\rangle.
\eea 
Note that $\psi$ can be expressed in terms of $\hat{Z}$ in a very complicated way so $\psi$ commutes with $\hat{Z}$. Then \eqref{inz2} can be rewritten in a very suggestive way
\bea \label{inz2a}
Z_\alpha^2=\langle \psi_\alpha^2 \hat{Z}^2\rangle=\langle \hat{Z}^2\rangle+\langle \psi_\alpha^2 \hat{Z}^2\rangle_c,
\eea 
where we have assumed that $\alpha$-state is normalized $\langle \psi_\alpha^2\rangle=1$. This rewriting is interesting because it separates out the self-averaged part $\langle \hat{Z}^2\rangle$ and non-self-averaged part $\langle \psi_\alpha^2 \hat{Z}^2\rangle_c$. 
In CGS model, since the eigenvalue of $\hat{Z}$ is continuous and supported on $\mathbb{R}$ so that we can express $\psi_\alpha$ in terms of $\hat{Z}$ schematically as
\bea 
&& \psi_\alpha =\delta(\hat{Z}-Z_\alpha)=\int \frac{\d k}{2\pi}e^{\im k(\hat{Z}-Z_\alpha)},
\eea  
thus 
\bea 
&&\langle Z^2 \psi_\alpha\rangle=\int \frac{\d k\, e^{-\im k Z_\alpha}}{2\pi} \langle Z^2 e^{\im k Z}\rangle\, \\
&&\rightarrow Z_\alpha^2=\int \frac{\d k\, e^{-\im k Z_\alpha}}{2\pi} \frac{\langle Z^2 e^{\im k Z}\rangle}{\langle \psi_\alpha \rangle}. \label{rm2}
\eea 
Noting that $\langle \psi_\alpha \rangle=P(Z_\alpha)$, where $P(Z)$ is the PDF of $Z$, we find that \eqref{rm2} coincides with the trick used in \cite{Peng:2022pfa} and \cite{Mukhametzhanov:2021hdi} of rewriting $Z_\alpha^n$ as a formal average 
\bea 
Z_\alpha^n=\int \d Z \delta(Z-Z_\alpha)\frac{Z^n P(Z)}{P(Z_\alpha)}=\int \frac{\d k}{2\pi}\frac{e^{-\im k Z_{\alpha}}}{P(Z_\alpha)}\langle Z^n e^{\im k Z}\rangle. \label{trick}
\eea 
From which we can derived some useful approximation formula  $Z_\alpha^n\approx \langle Z_\alpha^n\rangle+ \Phi$, where $\langle Z_\alpha^n\rangle $ and $\Phi$ are respectively recognized as the wormhole and half-wormhole contributions as shown in \cite{Peng:2022pfa,Mukhametzhanov:2021hdi}. So we can think of this trick as a refinement of the factorization proposal of \cite{Saad:2021uzi}.  
 We will elaborate this below.\par

\subsection{Half-Wormhole in CGS-like model}
In the CGS model, because $Z$ satisfies the Gaussian distribution there is a more concrete expression for the half wormhole saddle as shown in \cite{Saad:2021uzi}. The key point is the fact that when $Z$ is Gaussian, it can be thought of as  the position operator of a simple harmonic oscillator so there exists a natural orthogonal basis, the number basis $\{n\}$ which is called the $n$-baby universe basis in the context of the gravity model. If we insert the complete basis $\sum_i|i\rangle \langle i|$ into \eqref{trick} we can get\footnote{Note that our convention is $Z=\mu+t(a+a^\dagger)$}
\bea 
Z_\alpha^n&=&\int \d Z \delta(Z-Z_\alpha)\frac{Z^n P(Z)}{P(Z_\alpha)}=\int \frac{\d k}{2\pi}\frac{e^{-\im k Z_{\alpha}}}{P(Z_\alpha)}\langle Z^n e^{\im k Z}\rangle \\
&=&\int \frac{\d k}{2\pi}\frac{e^{-\im k Z_{\alpha}}}{P(Z_\alpha)}\sum_{i=0}^n\langle Z^n|i\rangle\langle i| e^{\im k Z}\rangle\\
&=&\int \frac{\d k}{2\pi}\frac{e^{-\im k Z_{\alpha}}}{P(Z_\alpha)}\sum_{i=0}^n{n\choose i}\langle Z^{n-i}\rangle \langle Z^i|i\rangle \langle i| e^{\im k Z}\rangle \\
&=&\int \frac{\d k}{2\pi}\frac{e^{-\im k Z_{\alpha}}}{P(Z_\alpha)}\sum_{i=0}^n{n\choose i}\langle Z^{n-i}\rangle \sqrt{i!} t^i\langle i| e^{\im k Z}\rangle\\
&=&\int \frac{\d k}{2\pi}\frac{e^{-\im k Z_{\alpha}}}{P(Z_\alpha)}\sum_{i=0}^n{n\choose i}\langle Z^{i}\rangle  \langle (a t)^{n-i} e^{\im k Z}\rangle\equiv \sum_{i=0}^n{n\choose i}\mu_i \theta^{(n-i)},\label{hwh}
\eea 
where 
\bea 
\theta^{(n-i)}= \int \frac{\d k}{2\pi}\frac{e^{-\im k Z_{\alpha}}}{P(Z_\alpha)}\langle e^{\im k Z}\rangle \frac{\langle (at)^{n-i}e^{\im k Z}\rangle}{\langle e^{\im k Z}\rangle },\quad \frac{\langle (at)^{n-i}e^{\im k Z}\rangle}{\langle e^{\im k Z}\rangle }\equiv \phi^c_{n-i}. 
\eea 
Note that
\bea 
&&\frac{\langle (at)^{n-i}e^{\im k Z}\rangle}{\langle e^{\im k Z}\rangle }\equiv \phi^c_{n-i}=(\im k t^2)^{n-i}=(\phi^c_1)^{n-i},\\
&&\theta^{(i)}= \int \frac{\d k}{2\pi}\frac{e^{-\im k Z_{\alpha}}}{P(Z_\alpha)}\langle e^{\im k Z}\rangle (\im k t^2)^{i}=\frac{(-t^2\partial_{Z_\alpha})^iP(Z_\alpha)}{P(Z_\alpha)},\label{detheta}
\eea 
then \eqref{hwh} coincides with results in \cite{Peng:2022pfa}. So we confirm the result that within the Gaussian approximation (only keep the first two cumulants), $Z^n_\alpha$ can be decomposed as \eqref{hwh} and it suggests that $\theta_i$'s are the convenient building blocks of possible half-wormhole saddles. Some examples of the decomposition \eqref{hwh} are \footnote{$\theta^{(i)}$ is simply the (unnormalized) Hermite polynomial.} 

\bea \label{de}
&&Z_\alpha^1=\theta^{(1)}+\langle Z\rangle,\label{de1}\\
&&Z_\alpha^2=\theta^{(2)}+2\langle Z\rangle\theta^{(1)}+\langle Z^2\rangle,\label{de2}\\
&&Z_\alpha^3=\theta^{(3)}+3\langle Z\rangle\theta^{(2)}+3\langle Z^2\rangle \theta^{(1)}+\langle Z^3\rangle,\label{de3}\\
&&Z_\alpha^4=\theta^{(4)}+4\langle Z\rangle \theta^{(3)}+6\langle Z^2\rangle\theta^{(2)}+4\langle Z^3\rangle \theta^{(1)}+\langle Z^4\rangle,\label{de4}
\eea 
with
\bea 
&&\theta^{(1)}=-\mu+Z_{\alpha},\quad \theta^{(2)}=(\mu-Z_\alpha)^2-t^2={\theta^{(1)}}^2-t^2,\\
&& \theta^{(3)}=-(\mu-Z_\alpha)^3+3(\mu-Z_\alpha)t^2={\theta^{(1)}}^3-3t^2\theta^{(1)},\\
&&\theta^{(4)}=3t^4-6t^2(\mu-Z_\alpha)^2+(\mu-Z_\alpha)^4={\theta^{(1)}}^4-6t^2{\theta^{(1)}}^2+3t^4.
\eea 
In general we have
\bea 
\theta^{(i)}= \int \frac{\d k}{2\pi}\frac{e^{-\im k Z_{\alpha}}}{P(Z_\alpha)}\langle e^{\im k Z}\rangle (\im k t^2)^{i}&=& \int \frac{\d k}{\sqrt{2\pi/t^2 }}e^{-\frac{(k-(\im (\mu-Z_\alpha)/t^2))^2}{2/t^2}}(\im k t^2)^i,
\eea 
so $\theta^{(i)} /(\im t^2)^i$ is the $i$-th moment of ``Gaussian distribution" $\mathcal{N}(\im (\mu-Z_\alpha)/t^2,1/t^2)$ and the generating function is
\bea  
  \langle e^{u k}\rangle_{k}=e^{\frac{\im u(\mu-Z_\alpha)}{t^{2}}+\frac{u^2}{2 t^2}}.
\label{thetagene}
\eea 
Considering the following ensemble average
\bea \label{kk}
\Big\langle\langle e^{u_1 k_1}\rangle_{k_1}\langle e^{u_2 k_2}\rangle_{k_2}\Big\rangle_{Z_\alpha}=e^{-\frac{u_1u_2}{t^2}},
\eea 
and expanding both sides into Taylor series of $u_1$ and $u_2$ one can find
\bea \label{orthogonal}
&&\langle \theta^{(i)}\theta^{(j)}\rangle_{Z_\alpha}=i!t^{2i}\delta_{ij}.
\eea 
Due to this orthogonal condition we can directly tell which sector in the decomposition of $Z_\alpha^n$ is dominant by computing $\langle Z^n_\alpha Z^n_\alpha\rangle$
\bea 
&&Z_\alpha^n=\sum_i c_i \theta^{(i)},\quad \langle Z_\alpha^nZ_\alpha^n\rangle=\sum_i c_i^2 i! t^{2i}.
\eea 
In CGS model, since there is only a single random variable $Z$ so it does not admit any approximation related to large $N$ or small $G_N$. Therefore the  wormhole or half-wormhole are not true saddles in the usual sense. To breath life into them we should consider a model with a large number $N$ of random variables such as random matrix theory or SYK model which can be described by certain semi-classical collective variables like the $G,\Sigma$ in SYK, which potentially have a dual gravity description. However we find that it is illustrative to firstly apply the factorization proposal to some simple statistical models as we did in \cite{Peng:2022pfa}.

\subsection{Statistical model}
 Let us consider a function $Y(X_i)$ of a large number $N$ independent random variables $X_i$. Assuming that $X_i$'s are drawn from the Gaussian distribution then we have the decomposition
 \bea \label{yn}
Y^n&=&\frac{1}{(2\pi)^N}\int \prod_i \(\d k_i \frac{e^{-\im  k_i X_i}}{P(X_i)}\)  \langle e^{\im \sum_ik_i x_i}\rangle \sum_{n_1,\dots,n_N}\langle Y^n|n_1,\dots,n_N\rangle\frac{\langle n_1,\dots,n_N|e^{\im \sum_ik_i x_i}\rangle}{ \langle e^{\im \sum_ik_i x_i}\rangle} \nonumber\\
&=&\sum_{k=\sum_i n_i}\Gamma_{k},
\eea 
where $\G_k$ denotes different sectors, in particular $\G_0=\langle Y^n\rangle$. This kind of model can be also thought of as the CGS model with species \cite{Saad:2021uzi}. 
\subsubsection{Simple observables}
The simplest operator is 
\bea \label{clt}
Y=\sum_{i=1}^N X_i.
\eea 
Apparently for $n=1$ there are only two sectors 
\bea 
&&Y=\sum_{i=1}^N \mu+\sum_{i=1}^N\theta_i^{(1)}=\Theta_0+\Theta_1,\quad \langle \Theta_0^2\rangle=N^2\mu^2,\quad \langle\Theta_1^2\rangle=Nt^2,
\eea 
and for $n=2$ there are three sectors
\bea 
&&Y^2=\Phi_0+\Phi_1+\Phi_2,\\
&&\Phi_0=\langle Y^2\rangle,\quad \Phi_1=2N\mu \sum_i\theta_i^{(1)},\\
&&\Phi_2=\sum_{ij}\(\theta_i^{(1)}\theta_j^{(1)}+\delta_{ij}(\theta_i^{(2)}-{\theta_i^{(1)}}^2)\)=\sum_{ij}\(\theta_i^{(1)}\theta_j^{(1)}-\delta_{ij}t^2\).
\eea 
In general the parameters $\mu$ and $t^2$ are $N$ independent therefore $Y^n$ is self-averaged $Y^n\approx \langle Y^n\rangle$ in the large $N$ limit. This is also true even $X_i$ are not Gaussian because of the central limit theorem. But we also know in the literature that in order to have well-defined semi-classical approximation, the parameters $\mu$ and $t^2$ should depend on $N$ in a certain way like in SYK model. Interestingly in this case if $t^2\sim \mu^2 N$, the self-averaged part and non-self-averaged part are comparable and we should keep them both. This is exactly what we have encountered in the 0-SYK model. But a crucial difference is that for this simple choice of observables, all the non-self-averaged sectors are also comparable so it is not fair to call any of them the half-wormhole saddle and to restore factorization we have to include all the non-self-averaged sectors. The extremal case is $t^2>>\mu^2 N$. In this limit we find that the sector with highest level  dominates. For example,
\bea 
&&Y\approx \Theta_1,\quad Y^2\approx \langle Y^2\rangle+\Phi_2,\label{z2h}\\
&&\langle \Theta_1^2\rangle\approx \langle Y^2\rangle,\quad \langle \Phi_2^2\rangle\approx 2\langle Y^2\rangle^2,
\eea 
then it is reasonable to identify $\Theta_1$ with half-wormhole and identify $\Phi_2$ with the 2-linked half-wormhole. Similarly we can introduce $n$-linked half-wormholes. For example, in this extremal case, we can approximate $Y^3$ with 
\bea 
&&Y^3\approx 3\langle Y^2\rangle \Theta_1+\Lambda_3,\label{simpley3}\\
&&\Lambda_3=\sum_{i\neq j\neq k}\(\theta_i^{(1)}\theta_j^{(1)}\theta_k^{(1)}\)+3\sum_{i\neq j}\theta_i^{(2)}\theta_j^{(1)}+\sum_i\theta_i^{(3)},\\
&&\quad ~=\sum_{i,j,k}\(\theta_i^{(1)}\theta_j^{(1)}\theta_k^{(1)}\)-3t^2N\sum_{i}\theta_i^{(1)},
\eea 
where the sector $\Lambda_3$ should describe the 3-linked half-wormhole. We will consider a similar construction in the 0-SYK model. 
\subsubsection{Exponential observables}
In the Random Matrix Theory or quantum mechanics, the most relevant observable is the exponential operator $\Tr(e^{\beta H})$ since it relates to the partition function. So it may be interesting to consider a similar exponential operator 
\bea\label{expy}
Y=\sum_i e^{\beta X_i}, 
\eea 
in the toy statistical model. By a Taylor expansion of the exponential operator we find the following decomposition
\bea 
e^{\beta X}=\langle e^{\beta X}\rangle \sum_{k}\frac{\beta^k\theta^{(k)}}{k!},\quad \theta^{(0)}\equiv 1,
\eea 
thus
\bea 
&&Y=\sum_k \Theta_k,\quad \Theta_k=e^{\mu\beta+\frac{\beta^2t^2}{2}}\sum_i \frac{\beta^{k}\theta_i^{(k)}}{k!},\label{ye1}\\
&& \langle \Theta_k^2\rangle=N e^{2\mu\beta+{t^2\beta^2}}\frac{(\beta t)^{2k}}{k!},\quad \frac{\langle \Theta_k^2\rangle}{\langle Y^2\rangle}=e^{-t^2\beta^2}\frac{(\beta t)^{2k}}{k!}\equiv r_k. \label{nspois}
\eea 
\begin{figure}[hb]
\centering
\subcaptionbox{}{ \includegraphics[scale=0.4]{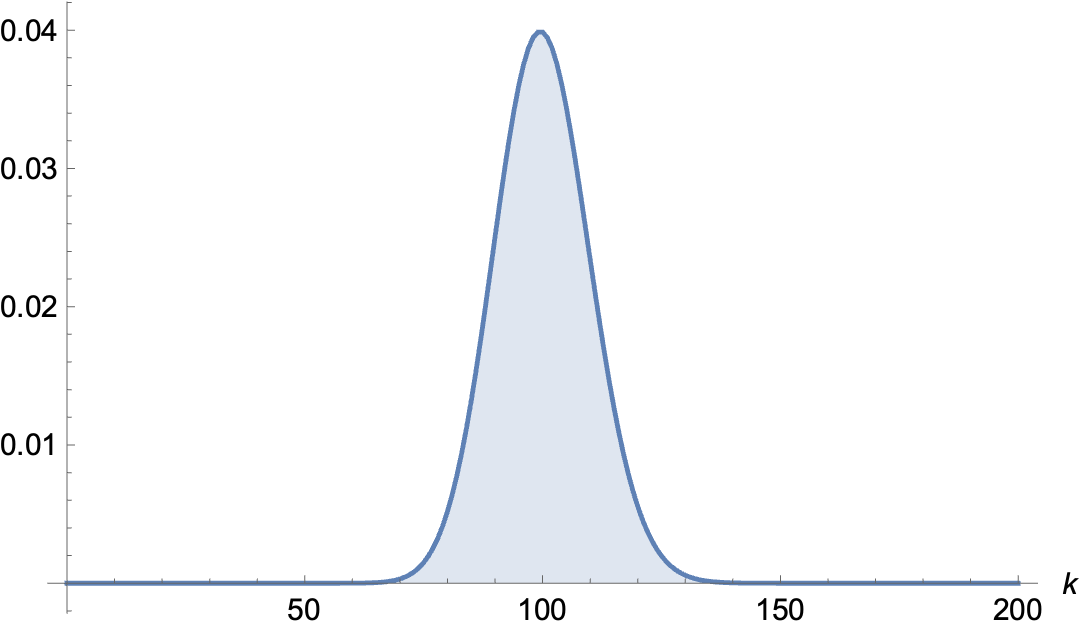}}
    \hfill
    \subcaptionbox{}{\includegraphics[scale=0.4]{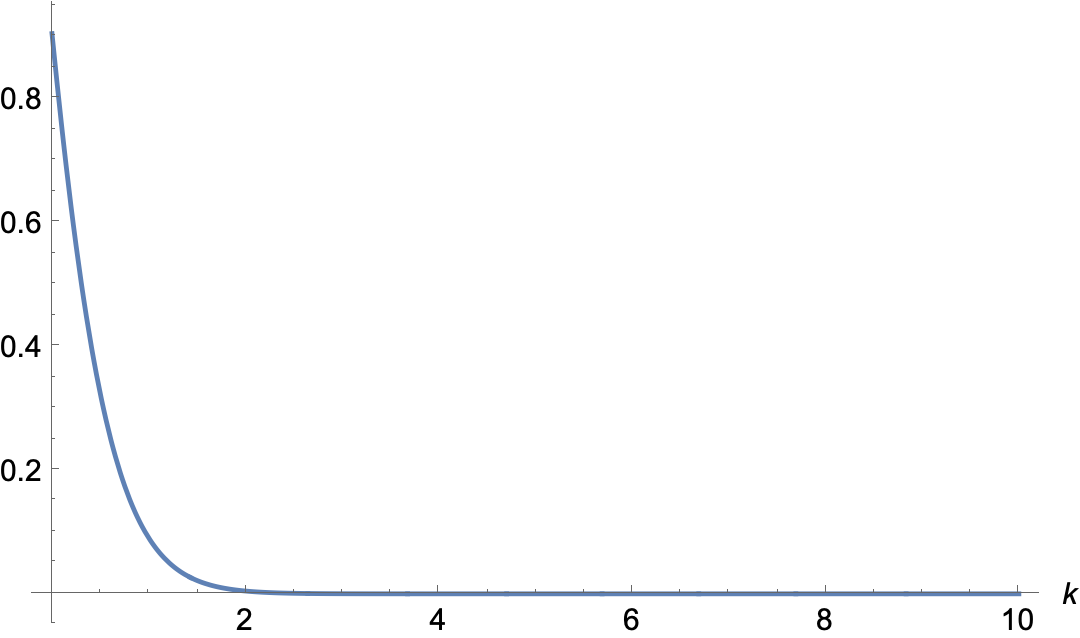}}
  \caption{Poisson distribution \eqref{nspois}. (a) Poisson distribution with $\beta^2 t^2=100$, (b) Poisson distribution with $\beta^2 t^2=0.01$.}
\end{figure}
Interestingly the ratio $r_k$ follows the Poisson distribution Pois$(\beta^{2} t^{2})$. When $\beta t <<1$  the dominant sector is $\Theta_0$ while for $\beta t>>1$ the Poisson distribution approaches Gaussian distribution $N(\beta^2 t^2,\beta^2 t^2)$ so we have to include all the sectors in the peak $k\in (\beta^2 t^2-{\beta t},\beta^2 t^2+ {\beta t})$ to have a good approximation.  
We can decompose $Y^2$ in a similar way
\bea 
&& Y^2=\sum_k \Phi_k,\label{y2kk}\\ 
&&\Phi_k=e^{2\mu\beta+\beta^2 t^2 }\beta^{k}\sum_{i\neq j}\sum_n\frac{\theta_i^{(n)}}{n!}\frac{\theta_j^{(k-n)}}{(k-n)!}+e^{2\mu\beta+2\beta^2 t^2 }(2\beta)^{k}\sum_i \frac{\theta_i^{(k)}}{k!},\\
&&\frac{\langle \Phi_k^2\rangle}{\langle Y^4\rangle}=e^{-2\beta^2 t^2}\frac{(2\beta^2t^2)^k}{k!}\frac{2(N-1)+2^k e^{2\beta^2 t^2}+4(N-1)e^{\beta^2t^2}}{2(N-1)+e^{4\beta^2t^2}+4(N-1)e^{\beta^2t^2}}
.\label{y2p}
\eea 
The behavior is similar. When $\beta t<<1$, the dominant sector is the self-averaged sector $\Phi_0$. When $2\beta^2 t^2> \log N$  \eqref{y2p} approaches the Gaussian $N(4\beta^2 t^2,4\beta^2t^2)$. On the other hand, when $1<<2\beta^2 t^2<< \log N$  \eqref{y2p} approaches the Gaussian $N(2\beta^2 t^2,2\beta^2t^2)$. In the end when $2\beta^2 t^2\sim \log N$, \eqref{y2p} will have two comparable peaks. 
\begin{figure}[hb]
\centering
  \includegraphics[scale=0.4]{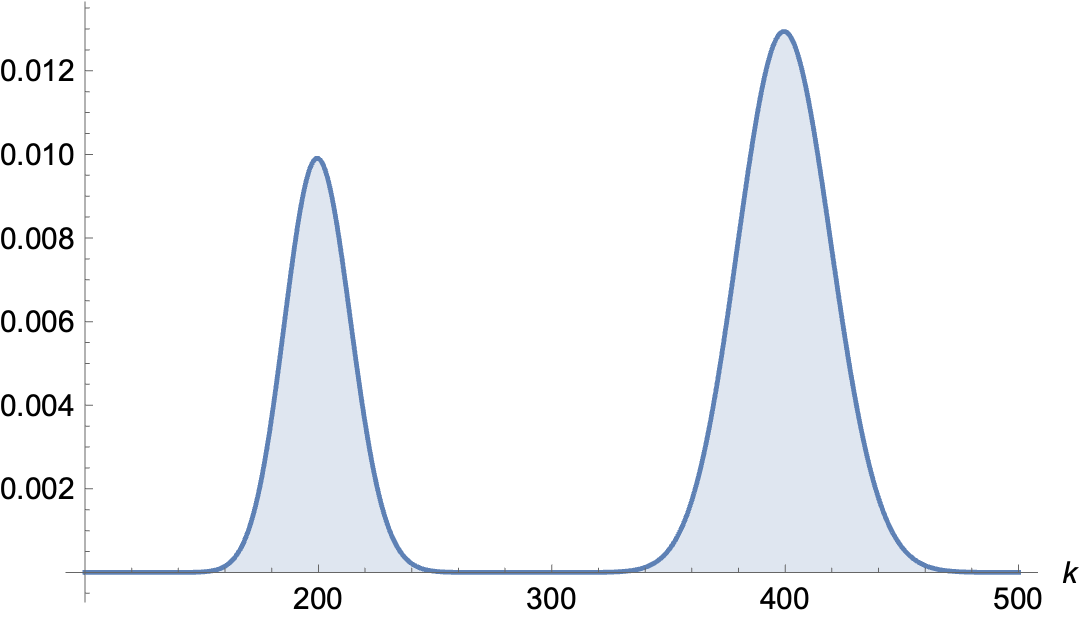}
  \caption{Plot of \eqref{y2p} when they are two comparable peaks. $\log N=298,\quad \beta t=10$.}
\end{figure}
However  the half-wormhole ansatz proposed in \cite{Mukhametzhanov:2021hdi,Peng:2022pfa} which can be written as
\bea 
&&\Phi=\sum_{k=0}^\infty \phi_k,\\
&&\phi_k=\Phi_k+(e^{-\beta^2 t^2 }-1)e^{2\mu\beta+2\beta^2 t^2 }(2\beta)^{k}\sum_i \frac{\theta_i^{(k)}}{k!},
\eea 
only works for small value of $\beta t$. 

To summarize our proposal, by introducing the basis $\{\theta_i\}$ which is the generalization of $n$-baby universe basis \cite{Saad:2021uzi} we can decompose the observables or partition functions into a single self-averaged sector and many non-self-averaged sectors. These sectors are independent in the sense of \eqref{orthogonal}. The contributions from each sector have interesting statistics: in the large $N$ limit leading contributing sectors may condense to peaks. This condensation is a signal that the observable potentially has a bulk description (or semi-classical description) in the large $N$ limit. If the self-averaged sector survives then it means the observable is approximately self-averaging. The surviving non-self-averaged sectors in the large $N$ limit are naturally interpreted as the ($n$-linked) half-wormholes which are the results of sector condensation. In the extremal case, only one non-self-averaging survives reminiscing the famous Bose-Einstein condensation.   

\subsection{0-SYK model}
\label{section0syk}
In this section we apply our proposal to the 0-SYK model which has the ``action"
\bea \label{z0}
z=\int \d^N \psi \exp (\im ^{q/2}\sum J_{i_1\dots i_q}\psi_{i_1\dots i_q})\,,
\eea
where $\psi_{i_1\dots i_q}=\psi_{a_1}\psi_{a_2}\dots \psi_{a_q}$ and $\psi_i$ are Grassmann numbers. 
The random couplings $J_{i_1\dots i_q}$ is drawn from a Gaussian distribution 
\bea 
\langle J_{i_1\dots i_q}\rangle=u,\quad \langle J_{i_1\dots i_q}J_{j_1\dots j_q}\rangle=t^2\delta_{i_1j_1}\dots \delta_{i_qj_q},\quad t^2=\tau^2\frac{(q-1)!}{N^{q-1}}\ ,\label{gauss}
\eea 
where we found in \cite{Peng:2022pfa} in order to have a semi-classical description $u$  should also have a proper dependence
\bea \label{mun}
u=(-\im)^{q/2}\mu\frac{(q/2-1)!}{2 N^{q/2-1}}.
\eea 
We sometimes use the collective indies $A,B$ to simplify the notation
\bea 
A=\{a_1<\dots < a_q\}\,,\qquad J_A\psi_A\equiv J_{a_1\dots a_q}\psi_{a_1\dots a_q}\ .
\eea 
Integrating out the Grassmann numbers directly gives \footnote{Here we choose the measure of Grassmann integral to be $\int d^N \psi \psi_{1\dots N}=\im^{-N/2}$.}:
\bea 
z=\int \d^N \psi \exp(\im ^{q/2}J_A \psi_A)=\sum'_{A_1<\dots<A_p} \text{sgn}(A)J_{A_1}\dots J_{A_p}\,,\quad p=N/q\,,\label{pfa}
\eea 
where the expression \eqref{pfa} is nothing but the hyperpfaffian $\text{Pf}(J)$.  According to \eqref{yn}, we can similarly decompose it as
\bea 
&&z=\sum_i\Theta_i,\quad \Theta_0=\langle z\rangle,\\
&&\Theta_k=u^{p-k}\sum'_{I_1<\dots <I_{p-k}}\text{Pf}(\theta^{(1)(I_1,\dots,I_{p-k})}_A),\label{thek}
\eea 
where the tensor $\theta^{(1)(I_1,\dots,I_{p-k})}_A$ means that the index $A$ is not in the set $(I_1,\dots,I_{p-k})$. The expression \eqref{thek} can be derived by a combinatorial method used in \cite{Peng:2022pfa} or by using the $G,\Sigma$ trick as follows. First we expand $z$ into series of $\theta^{(1)}$
\bea 
z&=&\int \d^N\psi e^{\im^{q/2}\sum_A J_A\psi_A}=\int \d^N\psi e^{\im^{q/2}\sum_A u \psi_A}e^{\im^{q/2}\sum_A \theta_A^{(1)} \psi_A}\label{z1} \\
&=&\int \d^N\psi \sum_{k=0}\frac{(\im^{q/2}  \sum_A \theta^{(1)}_A\psi_A)^{k}}{k!}e^{\im^{q/2}u\sum_A  \psi_A},
\eea 
thus by matching the power of $\theta^{(1)}$ we get a integral expression of $\Theta_k$
\bea 
\Theta_k=\int \d^N\psi \frac{(\im^{q/2}  \sum_A \theta_A^{(1)}\psi_A)^{k}}{k!}e^{\im^{q/2}u\sum_A  \psi_A}.
\eea 
Next following \cite{Peng:2022pfa} we can introduce $G,\Sigma$ variables directly as
\bea 
G&=&\frac{1}{N}\sum_{i<j}\psi_i\psi_j,\\
z&=&\int \d^N\psi \int_{\mathbb{R}}\d G \int_{\im \mathbb{R}}\frac{\d\Sigma}{2\pi \im/N} e^{u \im^{q/2}\frac{N^{q/2}}{(q/2)! }G^{q/2}} e^{- N\Sigma G}e^{\im^{q/2}\sum_A \theta_A^{(1)}\psi_A}e^{\Sigma\sum_{i<j}\psi_i\psi_j}\\
&=&\int \d^N\psi \int_{\mathbb{R}}\d G \int_{\im \mathbb{R}}\frac{\d\Sigma}{2\pi \im/N} e^{u \im^{q/2}\frac{N^{q/2}}{(q/2)! }G^{q/2}} e^{- N\Sigma G} \sum_k\frac{(\im^{q/2}  \sum_A \theta_A^{(1)} \psi_A)^k}{k!}e^{\Sigma\sum_{i<j}\psi_i\psi_j},\nonumber \\
\eea 
and 
\bea 
\Theta_k &=&\int \d^N\psi \int_{\mathbb{R}}\d G \int_{\im \mathbb{R}}\frac{\d\Sigma}{2\pi \im/N} e^{u \im^{q/2}\frac{N^{q/2}}{(q/2)! }G^{q/2}} e^{- N\Sigma G} \frac{(\im^{q/2}  \sum_A \theta_A^{(1)} \psi_A)^k}{k!}e^{\Sigma\sum_{i<j}\psi_i\psi_j},\\
&=&\int \d^N\psi \int_{\mathbb{R}}\d G \int_{\im \mathbb{R}}\frac{\d\Sigma}{2\pi \im/N} e^{u \im^{q/2}\frac{N^{q/2}}{(q/2)! }G^{q/2}} e^{- N\Sigma G} \frac{(\im^{q/2}  \sum_A \theta_A^{(1)} \psi_A)^k}{k!}\frac{(q/2 !)^{p-k}\Sigma^{\frac{N-qk}{2}} (\sum_A\psi_A)^{p-k}}{(N/2-qk/2)!} \nonumber \\
&=&\int_{\mathbb{R}}\d G \int_{\im \mathbb{R}}\frac{\d\Sigma}{2\pi \im/N} e^{u \im^{q/2}\frac{N^{q/2}}{(q/2)! }G^{q/2}} e^{- N\Sigma G} (\im \Sigma)^{\frac{N-qk}{2}} \frac{(q/2 !)^{p-k}(p-k)!}{(N/2-qk/2)!}\times \nonumber\\
&\quad & \int \d^N \psi \sum_{A_1<\dots <A_k}\theta_{A_1}^{(1)}\dots \theta_{A_k}^{(1)}\psi_{A_1}\dots\psi_{A_k}\times \sum_{I_1<\dots< I_{p-k}}\psi_{I_1}\dots\psi_{I_{p-k}}\\
&=&\int_{\mathbb{R}}\d G \int_{\im \mathbb{R}}\frac{\d\Sigma}{2\pi \im/N} e^{u \im^{q/2}\frac{N^{q/2}}{(q/2)! }G^{q/2}} e^{- N\Sigma G} (\im \Sigma)^{\frac{N-qk}{2}} \frac{(p-k)!(q/2!)^{p-k}}{(N/2-qk/2)!}\times   \sum'_{I_1<\dots <I_{p-k}}\text{PF}(\theta^{(1)(I_1,\dots,I_{p-k})}_A),\nonumber \\
&=& u^{p-k}\sum'_{I_1<\dots <I_{p-k}}\text{PF}(\theta^{(1)(I_1,\dots,I_{p-k})}_A), 
\eea 
where the tensor $\theta^{(1)(I_1,\dots,I_{p-k})}_A$ means that the index $A$ is not in the set $(I_1,\dots,I_{p-k})$. 
 To figure out which one is dominant let us compute
\bea 
\langle z^2\rangle=\sum_i \langle \Theta_i\Theta_i\rangle.
\eea 
The expression of $\langle z^2\rangle$ is derived in \cite{Peng:2022pfa}
\bea \label{z2our}
\langle z^2\rangle=\sum_{k=0}^p c_k m_{p-k}^2t^{2k}u^{2p-2k}\equiv \sum_k z_2^{(k)}\,,
\eea 
where
\bea 
&&c_k=\frac{1}{k!}{ N\choose q }{ N-q\choose q }\dots {N-(k-1)q\choose q}=\frac{N!}{k! (q!)^k (N-kq)!}\,, \label{ck}\\
&&m_p=\frac{(pq/2)!}{p!((q/2)!)^p}.\label{mp}
\eea
By matching the power of $t^2$ we can identify 
\bea 
z_2^{(k)}=\langle \Theta_k\Theta_k\rangle=c_k m_{p-k}^2t^{2k}u^{2p-2k}.
\eea 
The coefficient is very involved so let us first consider  some simple cases. If $p=2$, then there are only three sectors
\bea 
&&z=\langle z\rangle+\Theta_1+\Theta_2,\\
&&z_{2}^{(0)}=\frac{\left(q!\right)^{2}}{4\left(\frac{q}{2}!\right)^{4}}u^{4},\quad z_{2}^{(1)}=\frac{(2q)!}{\left(q!\right)^{2}}u^{2}t^{2},\quad z_{2}^{(2)}=\frac{(2q)!}{2\left(q!\right)^{2}}t^{4}.
\eea 
Taking the large $N$ limit, we find
\bea 
z_2^{(1)}\sim \sqrt{N}\frac{t^2}{u^2}z_2^{(0)},\quad z_2^{(2)}\sim \sqrt{N}\frac{t^4}{u^4}z_2^{(0)},
\eea 
and 
\bea
\frac{t^2}{u^2}\approx \frac{1}{N}\frac{\tau^2}{4\mu^2} \frac{\frac{N}{2}!}{ (\frac{N}{4}!)^2} \sim \frac{\tau^2}{\mu^2}\frac{2^{N/2}}{N\sqrt{N}},
\eea 
which implies that
\bea 
z_2^{(1)}\sim \frac{2^{N/2}}{N} z_2^{(0)},\quad z_2^{(2)}\sim \frac{2^{N}}{N^2\sqrt{N}}z_2^{(0)}\sim \frac{2^{N/2}}{N\sqrt{N}}z_2^{(1)},
\eea 
so that we have the approximation
\bea 
z\sim \Theta_2.
\eea 
Similarly when $p=3$, we can find
\bea 
z=\langle z\rangle+\Theta_1+\Theta_2+\Theta_3,
\eea 
and
\bea 
&&z_2^{(3)}\sim \frac{t^2}{u^2}\frac{1}{3}z_2^{(2)},\quad z_2^{(2)}\sim \frac{t^2}{u^2}\sqrt{N}z_2^{(1)},\quad z_2^{(1)}\sim \frac{t^2}{u^2}\sqrt{N}z_2^{(0)} ,\\
&&\frac{t^2}{u^2}\sim \frac{2^{N/3}}{N\sqrt{N}},
\eea 
thus
\bea 
z\approx \Theta_3.
\eea 
This turns out be general: when $p<<N$ the dominant term is $\Theta_p$. Therefore, the self-averaged $\langle z\rangle$ will not survive. This behavior is same as we found in the simple statistical model in the regime when the cylinder amplitude is much larger than the disk amplitude.\par
\begin{figure}[hb]
\centering
\subcaptionbox{}{ \includegraphics[scale=0.4]{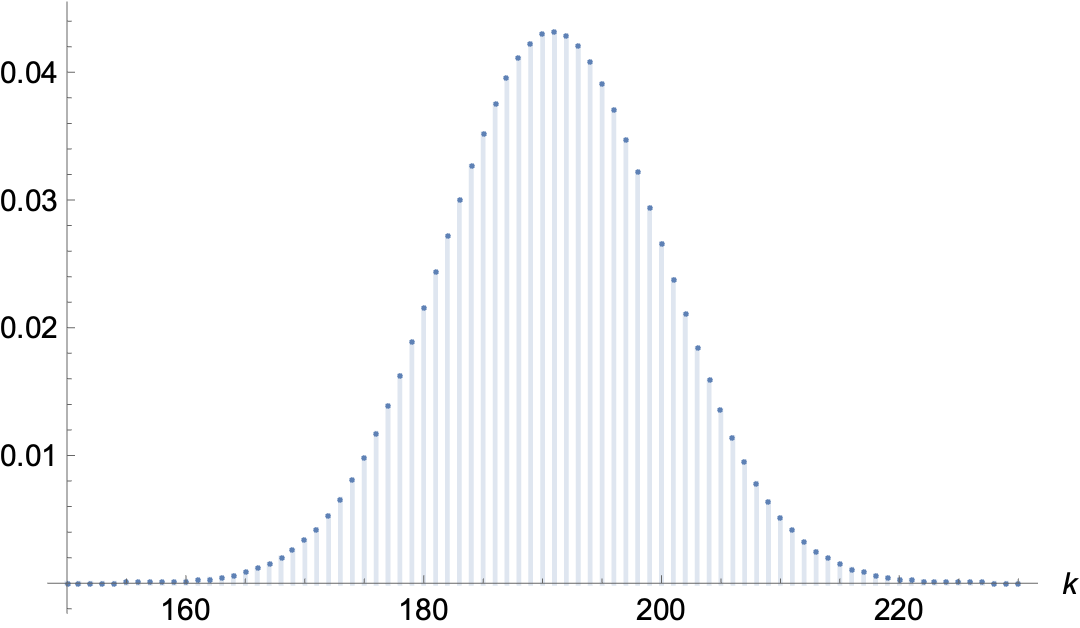}}
    \hfill
    \subcaptionbox{}{\includegraphics[scale=0.4]{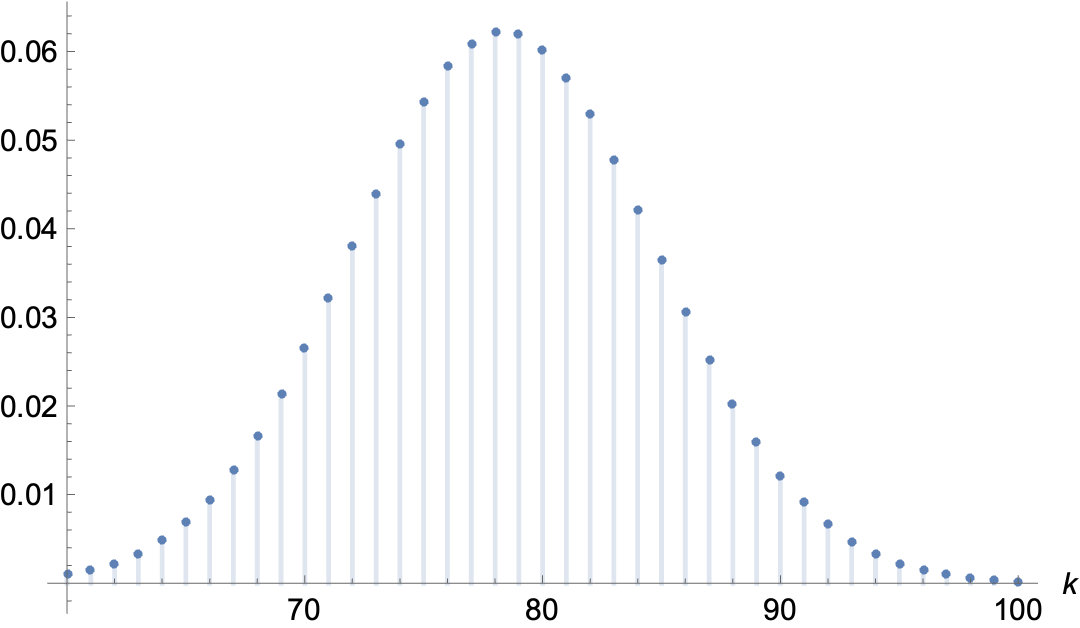}}
  \caption{The ratio $z^{(k)}_{2}/z_{2}^{(0)}$ in \eqref{coe}. (a) $N=1000,q=2$, the $y$ axis labels $\frac{s_k}{\sum_{i=0}^{p}s_i}$. The peak is of order $1/\sqrt{N}$. (b) $N=1000,q=4$, the $y$ axis labels $\frac{s_k}{\sum_{i=0}^{p}s_i}$. The peak is of order $1/\sqrt{N}$.}
  \label{pk}
\end{figure}

 
On the other hand, if $q<<N$ then
\bea 
\frac{t^2}{u^2}\sim \frac{\tau^2}{\mu^2}\sim\frac{1}{N},
\eea 
the situation is very different. As a simple demonstration let us consider the case of $q=2$
\bea 
&&z_2^{(p)}\sim N^{N/2}\frac{t^N}{u^N}z_2^{(0)}=\frac{1}{N^{N/2}}z_2^{(0)},\\
&&z_2^{(1)}\sim N^2\frac{t^2}{u^2}z_2^{(0)}=Nz_2^{(0)},\quad z_2^{(2)}\sim N^4\frac{t^4}{u^4}z_2^{(0)}=N^2z_2^{(0)},\\
&&z_2^{(3)}\sim N^6\frac{t^6}{u^6}z_2^{(0)}=N^3 z_2^{(0)},\dots,\\
&&z_2^{(k)}\sim s_kz_2^{(0)} ,\quad s_k=\frac{1}{2^k N^k}\frac{N!}{k!(N-2k)!}.\label{coe}
\eea 
The dominant term is neither  $\langle z\rangle $ nor $\Theta_p$ but some intermediate term $\Theta_k$ as argued in \cite{Peng:2022pfa}. With this detailed analysis we find that we should also include some ``sub-leading" sectors. The distribution of the  surviving sectors in the large $N$ limit has a peak centered at the ``dominant'' sector with a width roughly $\sqrt{N}$. One possible interpretation of this result is the surviving sectors are only approximate saddles or constrained saddles with some free parameters. Even though each approximate saddle contribution is as tiny as $1/\sqrt{N}$ but after integrating over the free parameters the total contribution is significant. Note that similar approximate saddles are also found for the spectral form factor in the SYK model \cite{Saad:2018bqo}.
 We plot the ratio $z_2^{(k)}/z_2^{(0)}$ as function of $k$ in Fig. \ref{pk}. With increasing $q$ or equivalently decreasing $p$, the peak moves to the left (small k) and becomes sharper and sharper. This is consistent with our analysis of limit of small $p$ where there is only one dominant saddle, $\Theta_p$.   
  So our result shows that the wormhole (actually disk in this case) does not persist but the half-wormhole appears.  
As we found in \cite{Peng:2022pfa} $\langle z^2\rangle $ can be computed by a trick of introducing the collective variables 
\bea 
G_{LR}=\frac{1}{N}\sum_i \psi_i^L \psi_i^R,\quad G_L=\frac{1}{N}\sum_{i<j}\psi_i^L \psi_j^L,\quad G_R=\frac{1}{N}\sum_{i<j}\psi_i^R \psi_j^R\,,
\eea 
and doing the path integral. The final expression is
\bea 
\langle z^2\rangle 
&=& \int_R \d^3 G_i \int_{\im \mathbb{R}}\d^3 \Sigma_i\, e^{\frac{N}{q}(\tau^2 G_{LR}^q+\mu G_L^{q/2}+\mu G_{R}^{q/2})- N(\Sigma_i G_i)}\frac{1}{2}\((\Sigma_{LR}+\im \sqrt{\S_L\S_R})^N+(\Sigma_{LR}-\im \sqrt{\S_L\S_R})^N\)\nonumber \\
&=&\int_R \d^3 G_i \int_{\im \mathbb{R}}\d^3 \Sigma_i\, \sum_{m=0}^{N/2} {N \choose 2m}( \Sigma_{LR})^{2m}(\im^2 \Sigma_L\Sigma_R)^{\frac{N}{2}-m} e^{\frac{N}{q}(\tau^2 G_{LR}^q+\mu G_L^{q/2}+\mu G_{R}^{q/2})}e^{- N(\Sigma_i G_i)}\nonumber\,, \label{exz2s}
\eea 
In \cite{Peng:2022pfa} we indeed find a new non-trivial saddle point whose saddle contribution is larger than the saddle contribution of the trivial disk saddle and wormhole saddle. The new non-trivial saddle should correspond to $\sum_{k}\langle \Theta_k^2\rangle$ with $k$ in the peak. The expression \eqref{exz2s}
 of $\langle z^2\rangle$ leads to a $G,\Sigma$ expression of each $z_2^{(k)}$
\bea 
\langle \Theta_k^2\rangle &=&z_2^{(k)}= {N \choose kq}\int_R \d^3 G_i \int_{\im \mathbb{R}}\d^3 \Sigma_i\, ( \Sigma_{LR})^{kq}(\im^2 \Sigma_L\Sigma_R)^{\frac{N-kq}{2}} e^{\frac{N}{q}(\tau^2 G_{LR}^q+\mu G_L^{q/2}+\mu G_{R}^{q/2})}e^{- N(\Sigma_i G_i)}\nonumber.\\ \label{gsk1}
\eea 
Actually we can derive a different $G,\Sigma$ expression from $\Theta_i$ directly in a more enlightening way.  Because $\psi_i$ are Grassmann numbers and $q$ is even then the exponential in \eqref{z1} factorizes
\bea \label{ee}
e^{\im^{q/2}\sum_A J_A\psi_A}=\prod_Ae^{\im^{q/2} J_A\psi_A}.
\eea 
Using Tyler expansion the definition of $\theta^{(i)}$ one can derive a useful identity
\bea \label{exp}
e^{\alpha X}=\langle e^{\alpha X}\rangle \sum_{i=0}^\infty \frac{\alpha^n}{n!}\theta^{(n)},
\eea 
where $X$ is the random variable. With the help of this identity and $\psi_A^2=0$, \eqref{ee} can be decomposed into
\bea 
\prod_Ae^{\im^{q/2} J_A\psi_A}&=&\langle e^{\im^{q/2}\sum_A J_A\psi_A}\rangle \(1+\sum_A \theta_A^{(1)}(\im^{q/2}\psi_A)+\frac{1}{2!}\sum_{A,B}\theta_A^{(1)}(\im^{q/2}\psi_A)\theta_B^{(1)}(\im^{q/2}\psi_B)+\dots \)\nonumber \\
&=&\langle e^{\im^{q/2}\sum_A J_A\psi_A}\rangle e^{\sum_A \theta_A^{(1)}\psi_A}.
\eea
Thus the we can express $\langle \Theta_k^2\rangle$ as
\bea 
&&\langle \Theta_k^2\rangle=\int \d^{2N}\psi^{L(R)}\langle e^{\im^{q/2}\sum_A J_A\psi^L_A}\rangle\langle e^{\im^{q/2}\sum_A J_A\psi^R_A}\rangle \frac{1}{k!^2}(\im^{q/2}\sum_A \psi_A^L \theta_A^{(1)} )^k(\im^{q/2}\sum_A \psi_A^R \theta_A^{(1)} )^k\nonumber \\
&&=\int \d^{2N}\psi^{L(R)}\langle e^{\im^{q/2}\sum_A J_A\psi^L_A}\rangle\langle e^{\im^{q/2}\sum_A J_A\psi^R_A}\rangle \frac{t^{2k}}{k!}(\sum_A\psi_A^L\psi_A^R)^k\\
&&=\int \d^{2N}\psi^{L(R)}\d G_{LR} \d\S_{LR}\, e^{\im^{q/2} u\sum_A(\psi_A^L+\psi_A^R)}e^{-N \S_{LR}(G_{LN}-\sum_i\psi_i^L\psi_i^R)}\frac{1}{k!}\(\frac{N\tau^2}{q}G_{LR}^q\)^k\nonumber\\
&&=\int_R \d^3 G_i \int_{\im \mathbb{R}}\d^3 \Sigma_i\, \frac{1}{2}\((\Sigma_{LR}+\im \sqrt{\S_L\S_R})^N+(\Sigma_{LR}-\im \sqrt{\S_L\S_R})^N\)\nonumber\\
&&\qquad \qquad \qquad e^{\frac{N}{q}(\mu G_L^{q/2}+\mu G_{R}^{q/2})}e^{- N(\Sigma_i G_i)}\frac{1}{k!}\(\frac{N\tau^2}{q}G_{LR}^q\)^k. \label{gsthe}
\eea 
The integral \eqref{gsthe} is not convergent but we can introduce the generating function
\bea 
&&F(v)=\int_R \d^3 G_i \int_{\im \mathbb{R}}\d^3 \Sigma_i\, \frac{1}{2}\((\Sigma_{LR}+\im \sqrt{\S_L\S_R})^N+(\Sigma_{LR}-\im \sqrt{\S_L\S_R})^N\)\nonumber\\
&&\qquad e^{\frac{N}{q}(v\tau^2 G_{LR}^q+\mu G_L^{q/2}+\mu G_{R}^{q/2})}e^{- N(\Sigma_i G_i)},
\eea 
which can be computed with a saddle point approximation and the $\langle \Theta_k^2\rangle$ is given by
\bea 
\langle \Theta_k^2\rangle=\frac{1}{k!}\frac{\d^k F(v)_{\text{saddle}}}{\d v^k}\Big|_{v=0}\, .
\eea 
As a simple test, we know that the exact result of $F(v)$ is just
\bea 
F(v)=\langle z^2\rangle_{t^2\rightarrow t^2 v}=\sum_k c_k m_{p-k}^2t^{2k}u^{2p-2k}v^k,
\eea 
which indeed leads to
\bea 
\langle \Theta_k^2\rangle=\frac{1}{k!}\frac{\d^k F(v)_{\text{saddle}}}{\d v^k}\Big|_{v=0}=c_k m_{p-k}^2t^{2k}u^{2p-2k}\,.
\eea 
\subsubsection{Half-wormhole in $z^2$}
To make the half-wormhole saddle manifest below we will  set $u=0$. In this case ``Bose-Einstein" condensation happens. As found in \cite{Saad:2021rcu}
 for the square of partition function $z^2$ the wormhole persists and there is only one dominant non-self-averaged sector. Applying \eqref{yn} directly leads to the decomposition
\bea 
&&z^2=\sum_i \Phi_{2i},
\eea 
with
\bea 
&&\Phi_0=\langle z^2\rangle= \sum'_{A_1(B_1)<\dots<A_p(B_p)}  \text{sgn}(A)\text{sgn}(B) t^2\delta_{A_1B_1} \dots \dots t^2 \delta_{A_p B_p},\\
&&\Phi_2=\sum_k  \sum'_{A_1(B_1)<\dots<A_p(B_p)}  \text{sgn}(A)\text{sgn}(B) t^2\delta_{A_1B_1} \dots (\theta_{A_k}^{(1)}\theta_{B_k}^{(1)}+\delta_{A_k B_k}(\theta_{A_k}^{(2)}-{\theta_{A_k}^{(1)}}^2))\dots t^2 \delta_{A_p B_p},\nonumber\\
&&\dots\\
&&\Phi_{2p}=\sum'_{A_1(B_1)<\dots<A_p(B_p)}  \text{sgn}(A)\text{sgn}(B)  (\theta_{A_1}^{(1)}\theta_{B_1}^{(1)}+\delta_{A_1 B_1}(\theta_{A_1}^{(2)}-{\theta_{A_1}^{(1)}}^2))\dots (\theta_{A_p}^{(1)}\theta_{B_p}^{(1)}+\delta_{A_p B_p}(\theta_{A_p}^{(2)}-{\theta_{A_p}^{(1)}}^2))\nonumber \\
\eea 
where $\Phi_{2p}$ is the half-wormhole saddle which is found in \cite{Saad:2021rcu,Mukhametzhanov:2021hdi} by noticing $\theta_A^{(1)}=J_A$ and $\theta_{A}^{(2)}-{\theta_{A}^{(1)}}^2=-t^2$. Actually the connection between the half-wormhole proposed in \cite{Saad:2021rcu} and factorization proposal introduced in \cite{Saad:2021uzi} has been pointed out in \cite{Peng:2021vhs}. A useful way to derive the expression of $\Phi_i$ is to use \eqref{exp}
 first 
\bea 
&&e^{\im^{q/2}\sum_A J_A(\psi_A^L+\psi_A^R)}=\prod_A e^{\im^{q/2} J_A(\psi_A^L+\psi_A^R)}\\
&&\quad =\langle e^{\im^{q/2}\sum_A J_A(\psi_A^L+\psi_A^R)} \rangle \prod_A(1+\im^{q/2}\theta_A^{(1)}(\psi_A^L+\psi_A^R)+\im^{q}\theta_A^{(2)}\psi_A^L\psi_A^R)\\
&&\quad=\langle e^{\im^{q/2}\sum_A J_A(\psi_A^L+\psi_A^R)} \rangle e^{\im^{q/2}\sum_{A}\theta_A^{(1)}(\psi_A^L+\psi_A^R)+\im^q\sum_A (\theta_A^{(2)}-{\theta_A^{(1)}}^2)\psi_A^L\psi_A^R}
\eea 
and then to substitute it into the integral form of $z^2$
\bea 
z^2&=&\int \d^{2N}\psi^{L(R)} \langle e^{\im^{q/2}\sum_A J_A(\psi_A^L+\psi_A^R)} \rangle e^{\im^{q/2}\sum_{A}\theta_A^{(1)}(\psi_A^L+\psi_A^R)+\im^q\sum_A (\theta_A^{(2)}-{\theta_A^{(1)}}^2)\psi_A^L\psi_A^R}\label{rez2}\\
&=&\int \d^{2N}\psi^{L(R)}e^{\im^{q/2}\sum_{A}\theta_A^{(1)}(\psi_A^L+\psi_A^R)+\im^q\sum_A [(\theta_A^{(2)}-{\theta_A^{(1)}}^2)+t^2]\psi_A^L\psi_A^R}\\
&=&\im^N\sum_{k=0}^p\int \d^{2N}\psi^{L(R)} \frac{\(\sum_A [(\theta_A^{(2)}-{\theta_A^{(1)}}^2)+t^2]\psi_A^L\psi_A^R\)^k}{k!}\frac{(\sum_A \theta_A^{(1)}\psi_A^L)^{p-k}}{(p-k)!}\frac{(\sum_A \theta_A^{(1)}\psi_A^R)^{p-k}}{(p-k)!}\nonumber \\
&=&\sum'_{A(B)}\text{sgn}(A)\text{sgn}(B')\prod_{i} (\theta_{A_i}^{(1)}\theta_{B_i}^{(1)}+\delta_{A_i B_i}(\theta_{A_i}^{(2)}-{\theta_{A_i}^{(1)}}^2+t^2)).
\eea 
By matching the power of $t^2$ we can extract the expression of $\Phi_i$.
Note that the expressions of $\Phi_i$ have been derived in \cite{Mukhametzhanov:2021hdi} based on the proposal of \cite{Saad:2021rcu}. In \cite{Mukhametzhanov:2021hdi} the non-dominant sectors are derived as fluctuations of the dominant saddle $\Phi_{2p}$ with the help of introducing $G,\Sigma$ variables.
Because our derivation here does not rely on $G,\Sigma$ trick so it can be used to derive possible $n$-linked half-wormholes in $z^n$. First we notice that $\langle z^2\rangle^2=\langle \Phi_0^2\rangle $ is in the same order of $\langle z^4\rangle \approx \langle z^2\rangle^2$ as proved in \cite{Saad:2021rcu} so the wormhole saddle persists. To confirm that $\Phi_{2p}$ is the only dominant non-self-averaged saddle we only need to show
\bea 
\langle z^4\rangle\approx \langle \Phi_0^2\rangle+\langle \Phi_{2p}^2\rangle,
\eea 
which also has been proved in \cite{Saad:2021rcu,Mukhametzhanov:2021hdi}. Another benefit of the rewriting \eqref{rez2} is that we can introduce $G,\Sigma$ variable directly if needed because the appearance of $\langle e^{\im^{q/2}\sum_A J_A(\psi_A^L+\psi_A^R)} \rangle$ instead of introducing them ``by hand" by inserting an identity as proposed in \cite{Saad:2021rcu}.  As we argued in \cite{Peng:2022pfa} when $u\neq 0$,  $\Phi_{2p}$ will not be the dominant sector anymore. Instead there will be a package of  surviving non-self-averaged sectors.

 \subsubsection{Half-wormhole in $z^3$}
 As we argued in the statistical toy model, there should exist $n$-linked half-wormholes. For simplicity let us focus on 3-linked half-wormholes and $z^3$. Similar to \eqref{rez2}, $z^3$ can be rewritten as
 \bea 
 z^3&=&\int \d^{3N}\psi^i e^{\im^{q/2}\sum_A J_A(\psi_A^1+\psi_A^2+\psi_A^3)}\\
 &=& \int \d^{3N}\psi^i\langle e^{\im^{q/2}\sum_A J_A(\psi_A^1+\psi_A^2+\psi_A^3)}\rangle  e^{\im^{q/2}\sum_A \theta_A^{(1)}(\psi_A^1+\psi_A^2+\psi_A^3)}\times\nonumber \\
 && e^{\im^{q}\sum_A(\theta_{A}^{(2)}-{\theta_A^{(1)}}^2)(\psi_A^{1}\psi_A^{2}+\psi_A^{1}\psi_A^{3}+\psi_A^{2}\psi_A^{3})}e^{\im^{3q/2}\sum_A(\theta_A^{(3)}-3\theta_{A}^{(2)}{\theta_A^{(1)}}+2{\theta_A^{(1)}}^3)\psi_A^1\psi_A^2\psi_A^3},\\
 &=&\sum_{i=0}^{3p}\L_i.
 \eea 
 Again the expression of $\L_i$ can be extracted by matching the power of $t^2$.
 Since $\langle z^3\rangle=0$, so the self-averaged sector does not exist and $z^3$ is only dominated by non-self-averaged sectors which we expect are $\Lambda_{3p}$:
 \bea 
 \L_{3p}&=&\int \d^{3N}\psi^i e^{\im^{q/2}\sum_A \theta_A^{(1)}(\psi_A^1+\psi_A^2+\psi_A^3)}\times\nonumber \\
 && e^{\im^{q}\sum_A(\theta_{A}^{(2)}-{\theta_A^{(1)}}^2)(\psi_A^{1}\psi_A^{2}+\psi_A^{1}\psi_A^{3}+\psi_A^{2}\psi_A^{3})}e^{\im^{3q/2}\sum_A(\theta_A^{(3)}-3\theta_{A}^{(2)}{\theta_A^{(1)}}+2{\theta_A^{(1)}}^3)\psi_A^1\psi_A^2\psi_A^3},\\
 &=&\int \d^{3N}\psi^i e^{\im^{q/2}\sum_A J_A(\psi_A^1+\psi_A^2+\psi_A^3)}e^{-\im^{q}t^2\sum_A(\psi_A^{1}\psi_A^{2}+\psi_A^{1}\psi_A^{3}+\psi_A^{2}\psi_A^{3})},\label{lp3}
 \eea 
 and $\L_{p}$:
 \bea \label{lp1}
 \Lambda_{p}&=& \int \d^{3N}\psi^{1,2,3}\sum_{(i,j,k)=(1,2,3),(1,3,2),(2,3,1)}\( e^{\im^q t^2 \sum_A\psi_A^i\psi_A^j} e^{\im^{q/2}\sum_A \theta_A^{(k)}\psi_A^k}\)\\
 &=&3\langle z^2\rangle z,
 \eea 
 where we have substituted the explicit expressions of $\theta^{(i)}_A$. The term of triple product $\psi_A^1\psi_A^2\psi_A^3$ drops out in $\L_{3p}$ because of $J_A$ is Gaussian so that there is  no tri-linear interactions. From \eqref{lp3} and \eqref{lp1} it is obvious to show $\langle \L_{3p}\rangle=\langle \L_p\rangle=0$ as they should be. To confirm that they are  dominant let us compute $\langle \L_{3p}^2\rangle$ and $\langle \L_{p}^2\rangle$
 \bea 
 \langle \L_{3p}^2\rangle &=&\langle \L_{3p}^L\L_{3p}^R\rangle\\
 &=&\int  \d^{6N}\psi^{L_i(R_i)}e^{\im^q t^2\sum_A\(\sum_{i,j=1}^{3}\psi_A^{L_i}\psi_A^{R_j}\)}\approx 6\langle z^2\rangle^3, \\
\langle \L_{p}^2\rangle &= &\langle \L_{p}^L\L_{p}^R\rangle=9\langle z^2\rangle^3,
 \eea 
 which give
 \bea 
 \langle \L_{3p}^2\rangle+\langle \L_{p}^2\rangle\approx 15\langle z^2\rangle^3\approx \langle z^6\rangle.
 \eea 
 Therefore the approximation
 \bea 
 z^3\approx \L_p+\L_{3p},
 \eea 
 is the analogue of \eqref{simpley3}. We believe that this analogy persists for all other higher moments  $z^n$. Recall that $\theta^{(i)}$ can be thought of as moments thus it is reasonable to introduce the connected moments or the cumulants $\tilde{\theta}^{(i)}$ with those $z^3$ can be cast into
 \bea 
 z^3=\int \d^{3N}\psi^{i}\langle e^{\im^{q/2}\sum_A J_A(\psi_A^1+\psi_A^2+\psi_A^3)}\rangle \prod_k e^{\im^{kq/2}\sum_A \tilde{\theta}_A^{(k)}\frac{(\sum_i \psi_A^i)^k}{k!}}.
 \eea 
 In general, we expect
 \bea 
 z^n=\int \d^{nN}\psi^{i}\langle e^{\im^{q/2}\sum_A J_A(\sum_{i=1}^n\psi_A^i)}\rangle \prod_{k=1}^n e^{\im^{kq/2}\sum_A \tilde{\theta}_A^{(k)}\frac{(\sum_i \psi_A^i)^k}{k!}},
 \eea 
 which is simple to check for small $n$ by a direct calculation. Since $\theta^{(i)}$ is Gaussian so the only non-vanishing cumulants are $\tilde{\theta}^{(1)}$ and $\tilde{\theta}^{(2)}$ thus
 \bea \label{dzn}
 z^n=\int \d^{nN}\psi^{i}\langle e^{\im^{q/2}\sum_A J_A(\sum_{i=1}^n\psi_A^i)}\rangle e^{\im^{q/2}\sum_A \theta_A^{(1)}(\sum_i\psi_A^{i})}e^{\im^q\frac{1}{2}\sum_A\tilde{\theta}_A^{(2)}(\sum_i\psi_A^i)^2}.
 \eea 
 As a consistency check, substituting the explicit expressions $\theta^{(1)}_A=J_A$ and $\tilde{\theta}^{(2)}_A=-t^2$ into \eqref{dzn} leads to $z^n$ directly as it should be since \eqref{dzn} is nothing but a rewriting of $z^n$ in a convenient way of extracting contributions from different sectors and it is a direct generalization of the trick introduced in \cite{Saad:2021rcu}. In particular the highest level sector of $z^n$ can be expressed as
\bea\label{H} 
\Theta=\int \d^{nN}\psi^{i} e^{\im^{q/2}\sum_A \theta_A^{(1)}(\sum_i\psi_A^{i})}e^{\im^q\frac{1}{2}\sum_A\tilde{\theta}_A^{(2)}(\sum_i\psi_A^i)^2},
\eea 
which is expected to be one of the dominant non-self-averaged sector in the large $N$ limit.
\subsection{0+1 SYK model}
 Now let us apply our proposal to the 1-SYK model. The partition function is defined as
\bea 
z(\beta)=\int D\psi \, \exp \left\{-\int_0^\beta \d \tau\(\psi_i\partial_\tau \psi_i+\im^{q/2} J_A\psi_A\)\right\},
\eea 
with $J_A$'s satisfy \eqref{gauss}. We will assume that \eqref{dzn} is approximately valid at least semi-classically. In other words, the saddle point can be derived from \eqref{dzn}. The possible problem of \eqref{dzn} in one-dimensional SYK model is that the fermions are not Grassmann numbers but Majorana fermions. As a result, $\psi_A$ does not commute with $\psi_B$ if there are odd number common indexes in the collective indexes $A$ and $B$. Therefore \eqref{exp} is not exact anymore. The reason why we expect such subtlety is negligible in the large $N$ limit is because when we introduce standard $G,\Sigma$ variables in the SYK model we already ignore this fact and it is shown in \cite{Saad:2018bqo} this approximation is correct in the large $N$ limit. 
\subsubsection{Half-wormhole in $z$ and complex coupling}
First let us consider $z(\beta+\im T)$
\bea \label{zbt}
z(\beta+\im T)=\int D\psi \, e^{-\int_0^\beta \d \tau\(\psi_i\partial_\tau \psi_i\)+t^2\sum_A \mathcal{O}_A\mathcal{O}_A}e^{\sum_A \theta_A^{(1)}\mathcal{O}_A}e^{\sum_A \tilde{\theta}_A^{(2)}\mathcal{O}_A\mathcal{O}_A},
\eea 
where we have defined the operator
\bea 
\mathcal{O}_A(\beta+\im T)\equiv \im^{q/2}\int_0^{\beta+\im T} \d\tau_A \psi_A.
\eea 
The reason we consider $z(\beta+\im T)$ is that its square $\langle z(\beta+\im T) z(\beta-\im T)\rangle$  is the spectral form factor (SFF) which has universal behaviors for chaotic systems like SYK model and random matrix theories. When $T$ is small, SFF is self-averaged so it is dominated by disconnected piece $\langle z(\beta+\im T) z(\beta-\im T)\rangle\approx \langle z(\beta+\im T\rangle \langle z(\beta-\im T)\rangle$. Because the one point function  decays with respect to time and so is SFF. This decay region of SFF is called the slope. Because of the chaotic behavior SSF should not vanish in the late time. It will be the non-self-averaged sector dominates which are responsible for the ramp of the SFF. Therefore, in the ramp region we expect the approximation
\bea 
z\approx \Theta(\beta+\im T)&\equiv &\int D\psi \, e^{-\int_0^{\beta+\im T} \d \tau\psi_i\partial_\tau \psi_i}e^{\sum_A \theta_A^{(1)}\mathcal{O}_A}e^{\sum_A \tilde{\theta}_A^{(2)}\mathcal{O}_A\mathcal{O}_A}\\
&=&\int D\psi \, e^{-\int_0^{\beta+\im T} \d \tau\psi_i\partial_\tau \psi_i}e^{\sum_A \theta_A^{(1)}\mathcal{O}_A}e^{-\frac{t^2}{2}\sum_A \mathcal{O}_A\mathcal{O}_A},
\eea 
which is the analog of the highest level sector \eqref{H} in the 0d SYK model. It can also be written as $\langle e^{\sum_A \theta_A^{(1)}\mathcal{O}_A}\rangle_{\overline{SYK}}$, where $\overline{SYK}$ can be thought of as the anti-SYK model which is a SYK model but with an opposite  bi-linear coupling or it can be think of as a SYK model with purely imaginary random coupling $\im \tilde{J}_A$. 
The relation between factorization and complex couplings in SYK model was also proposed in \cite{Mukhametzhanov:2021hdi}. To confirm this approximation, let us compute
\bea 
\langle \Theta(\beta+\im T)\Theta(\beta-\im T)\rangle &=& \int D\psi^LD\psi^R e^{-\int_0^{\beta\pm \im T} \d \tau\psi_i^{L(R)}\partial_\tau \psi_i^{L(R)}} e^{-\frac{t^2}{2} (\mathcal{O}_A^L\mathcal{O}_A^L+\mathcal{O}_A^R\mathcal{O}_A^R)}\times\nonumber \\
&& \langle e^{\sum_A\theta_A^{(1)}\mathcal{O}_A^L}e^{\sum_A\theta_A^{(1)}\mathcal{O}_A^R}\rangle \\
&= & \int D\psi^LD\psi^R e^{-\int_0^{\beta\pm \im T} \d \tau\psi_i^{L(R)}\partial_\tau \psi_i^{L(R)}} e^{t^2 \sum_A\mathcal{O}_A^L\mathcal{O}_A^R},\label{th22}
\eea 
which  describes the wormhole saddle considering that we can introduce the $G_{LR}$ as 
\bea 
t^2\sum_A \mathcal{O}^L_A\mathcal{O}^R_A\approx \frac{t^2}{q!}\int \d\tau_L\int \d\tau_R (\sum_i\psi_i^L\psi_i^R)^q \equiv \frac{N\tau^2}{q}\int \d\tau_L\int \d\tau_R G_{LR}^q.
\eea
so the saddle point solution of \eqref{th22} is the same saddle point solution of $\langle z^2\rangle$ with $G_{LL}=G_{RR}=0$. Such solutions are found in \cite{Saad:2018bqo}. To be more precise, these solutions found in \cite{Saad:2018bqo} are time-dependent and only in the ramp region we have $G_{LL},G_{RR}\rightarrow 0$. This is why we stress that only in the ramp region our approximation is good. Away from this region, we have to include other sectors which can be obtained by the expansion \eqref{zbt} as
\bea 
\Theta_k &=& \int D\psi \, e^{-\int_0^\beta \d \tau\(\psi_i\partial_\tau \psi_i\)}e^{\sum_A \theta_A^{(1)}\mathcal{O}_A}e^{\sum_A \tilde{\theta}_A^{(2)}\mathcal{O}_A\mathcal{O}_A}\(\frac{(t^2\sum_A\mathcal{O}_A\mathcal{O}_A)^k}{k!}\),\\
&\approx &\frac{1}{k!}\left\langle e^{\sum_A \theta_A^{(1)}\mathcal{O}_A}\(\frac{N\tau^2}{q}G_{LL}^q\)^k \right\rangle_{\overline{SYK}}.
\eea 
 
\subsubsection{Half-wormhole in $z^2$ and factorization}
\begin{figure}[h]
\centering
  \includegraphics[width=3cm]{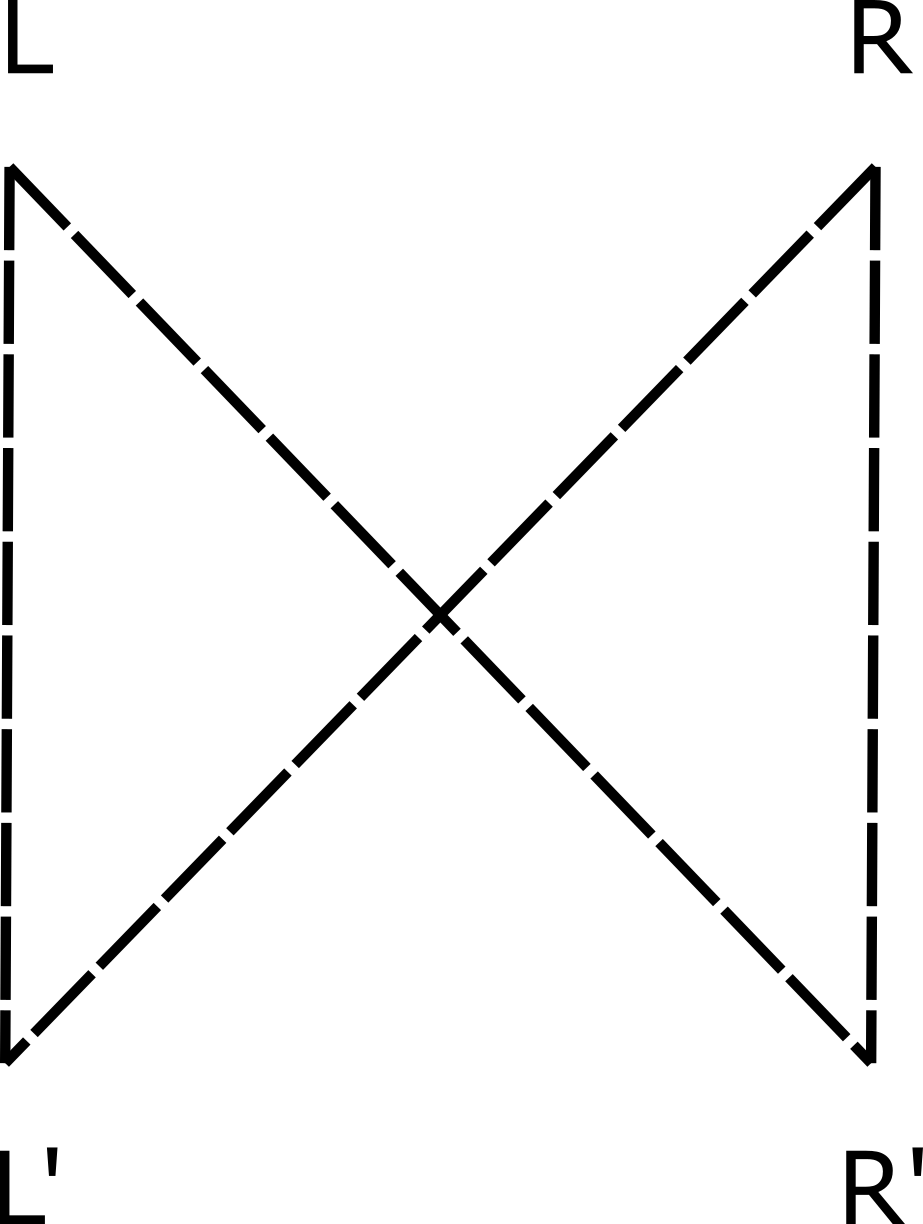} 
  \caption{The illustration of \eqref{nill}.}
  \label{hw2}
\end{figure}

Let us consider $z(\im T)z(-\im T)$ and apply our decomposition proposal \eqref{dzn}
\bea 
z(\im T)z(-\im T)= \int D\psi^{L(R)} \, e^{S_{SYK}^L+S_{SYK}^R}e^{\sum_A \theta_A^{(1)}(\mathcal{O}_A^L+\mathcal{O}_A^R)}e^{\frac{1}{2}\sum_A \tilde{\theta}_A^{(2)}(\mathcal{O}_A^L+\mathcal{O}_A^R)^2}.
\eea 
Motivated by the result of 0-SYK model, we expect that there is also a ramp region where the dominant non-self-averaged sector is given by the 2-linked half-wormhole\footnote{Note that we have normalized the fermionic integral such that $\int d\psi=0$ thus $\langle \Phi\rangle=0$.}
\bea 
\Phi&=&\int D\psi^{L(R)} \, e^{\sum_A \theta_A^{(1)}(\mathcal{O}_A^L+\mathcal{O}_A^R)}e^{\frac{1}{2}\sum_A \tilde{\theta}_A^{(2)}(\mathcal{O}_A^L+\mathcal{O}_A^R)^2}\\
&=&\int D\psi^{L(R)} \, e^{\sum_A \theta_A^{(1)}(\mathcal{O}_A^L+\mathcal{O}_A^R)}e^{-\frac{t^2}{2}\sum_A (\mathcal{O}_A^L+\mathcal{O}_A^R)^2}\\
&=&\int D\psi^{L(R)} \, e^{\sum_A \theta_A^{(1)}(\mathcal{O}_A^L+\mathcal{O}_A^R)}e^{-\frac{t^2}{2}\sum_A \(\mathcal{O}_A^L\mathcal{O}_A^L+\mathcal{O}_A^R\mathcal{O}_A^R+\mathcal{O}_A^L\mathcal{O}_A^R+\mathcal{O}_A^R\mathcal{O}_A^L\)} \label{1ph2}.
\eea 
Our proposal \eqref{1ph2} of the 2-linked half-wormhole is very close to the one 
proposed in \cite{Mukhametzhanov:2021hdi} which has two more bi-linear terms $\mathcal{O}_A^L\mathcal{O}_A^L+\mathcal{O}_A^R\mathcal{O}_A^R$ in the second exponent. It seems that our proposal is more proper 
considering that in $\langle \Phi^2\rangle$ there are only bi-linear correlations between $L(R)$ and $L'(R')$ 
\bea 
\langle \Phi^2\rangle=&&\int D\psi^LD\psi^R e^{-\int \d \tau\psi_i^{L(R)}\partial_\tau \psi_i^{L(R)}} \int D\psi^{L'} D\psi^{R'} e^{-\int \d \tau\psi_i^{L'(R')}\partial_\tau \psi_i^{L'(R')}}\nonumber \\
&&e^{t^2 \sum_A (\mathcal{O}_A^L\mathcal{O}_A^{L'}+\mathcal{O}_A^L\mathcal{O}_A^{R'}+\mathcal{O}_A^R\mathcal{O}_A^{L'}+\mathcal{O}_A^R\mathcal{O}_A^{R'})},
\label{nill}
\eea 
as shown in Fig. \ref{hw2}. Thus it implies the approximate factorization
\bea 
&&\text{Error}=z^2-\langle z^2\rangle-\Phi ,\\
&&\langle \text{Error}^2\rangle\approx \langle z^4\rangle-\langle z^2\rangle^2-2\langle \Phi z^2\rangle+\langle \Phi^2\rangle\approx (3-1-4+2)\langle z^2\rangle^2\approx 0,
\eea 
where we have assumed in the regime where the wormhole dominates the partition function $z$ approximates a Gaussian random variable. The bulk point of view of the factorization is also interesting. The insertion of $e^{\sum_A \theta_A^{(1)}(\sum_{i=1}^n\mathcal{O}^i_A)}$ can be thought of inserting spacetime branes in the gravity path integral and the opposite bi-linear coupling means the wormhole amplitudes connecting the branes are opposite to the usual spacetime wormhole amplitudes such that including all the effects of wormholes and branes factorization is achieved. In \cite{Blommaert:2021fob}, it is proposed that JT gravity can be factorized by inserting such spacetime branes.  

\subsection{Random Matrix Theory}
In this section, let us apply our proposal to the Random Matrix Theory: the GUE ensemble which can also be thought of the CGS model with End-Of-World (EOW) branes. The random matrix element  $H_{ij}$ is identified with a EOW brane $(\hat{\psi}_i,\hat{\psi}_j)$ in the notation of \cite{Marolf:2020xie} or the topological complex matter field $Z_{\psi^\dagger\psi}$ in the notation of \cite{Peng:2021vhs} with the restriction that the disk amplitude of $(\hat{\psi}_i,\hat{\psi}_i)$ vanishes i.e. $\langle (\hat{\psi}_i,\hat{\psi}_i)\rangle=0$. The equivalence between these two models can be understood as the following. The correlation functions of $H_{ij}$ are computed by the Wick contractions which exactly describe how to connect different EOW branes $(\hat{\psi}_i,\hat{\psi}_j)$ with spacetime wormholes in the Disk-Cylinder approximation. Therefore the correlation functions of the random matrix theory are equal to the gravity path integral as we have seen in the CGS model. In this theory, we are interested in the observable 
\bea 
z(\beta)=\Tr(e^{-\beta H})\, ,
\eea 
whose ensemble average is given by
\bea 
\langle z\rangle=\int \d{H}\, e^{-\frac{1}{2t^2} \Tr H^2}z\, ,
\eea 
where $t^2$ is usually taken to be $1/N$.
\subsubsection{Half-wormhole }
\label{matrixde}
First let us consider the non-self-averaged sector in $z$.   It is useful to study a simpler observable $\Tr H^n$ to get some intuitions about the non-self-averaged sector of matrix functions. For the random variable $H_{ij}$ we can not use the decomposition \eqref{yn} directly. One possible way of adapting to  \eqref{yn} is to rewrite $H_{ij}$ as a linear combination of the Gaussian random variables. However this rewriting is not very convenient. Alternatively, we can transfer the matrix integral into the integral over eigenvalues
\bea 
\mathcal{Z}(H)=\int \d He^{-\frac{1}{2t^2} \tr{H^2}}=\int \prod_i\d\lambda_i e^{-\frac{1}{2t^2}\sum_i\lambda_i^2}\Delta(\lambda)^2,
\eea 
where $\Delta(\lambda)$ is the Vandermonde determinant
\bea 
\Delta(\lambda)=\prod_{i<j}^L(\lambda_i-\lambda_j).
\eea 
Then the simple single-trace observable translates to
\bea 
\Tr H^n=\sum_i\lambda_i^n\,. 
\eea 
However those eigenvalues are not Gaussian random variables. As a result, even though we can still do the sector decomposition but the resulting different sectors are not orthogonal anymore. Although when the level is finite, we can obtain a new orthogonal basis by a direct diagonalization but it is still very cumbersome. We will make some preliminary analysis beyond Gaussian distribution in next section. Here we will take a similar approach as before. Considering the non-vanishing correlator $\langle H_{ij}H_{ji}\rangle=t^2$ we should define 
\bea 
\theta_{ij}^{(1)}=H_{ij},\quad \theta_{ij,ji}^{(2)}= \theta_{ij}^{(1)}\theta_{ji}^{(1)}-t^2,\label{pair}
\eea 
thus we have
\bea
H_{ij}H_{km}&=&\langle H_{ij}H_{km}\rangle+\theta_{ij}^{(1)}\theta_{km}^{(1)}+\delta_{jk}\delta_{im}(\theta^{(2)}_{ij,ji}-\theta_{ij}^{(1)}\theta_{ji}^{(1)})\\
&\equiv & \langle H_{ij}H_{km}\rangle+ [H_{ij}H_{km}]
\eea 
and 
\bea 
&&H_{ij}H_{kl}H_{mn}=\sum_{i=0}^3 \Theta_i,\quad \Theta_0=\langle H_{ij}H_{kl}H_{mn}\rangle,\quad \Theta_2=0,\\
&&\Theta_1=\theta_{ij}^{(1)}\langle H_{kl}H_{mn}\rangle+\theta_{kl}^{(1)}\langle H_{ij}H_{mn}\rangle+\theta_{mn}^{(1)}\langle H_{kl}H_{ij}\rangle,\\
&&\Theta_3=\begin{cases}
		\theta_{ij}^{(1)} \theta_{kl}^{(1)} \theta^{(1)}_{mn},\quad \text{no pairs like }  \eqref{pair} \\
		\theta_{ab,ba}^{(2)}\theta_{ji}^{(1)}=\theta_{ab}^{(1)}\theta_{ba}^{(1)}\theta_{ji}^{(1)}-t^2\theta_{ji}^{(1)},\quad \text{there is only one pair like }  \eqref{pair}\\
		\theta_{ab,ba,ab}^{(3)}=\theta_{ab}^{(1)}\theta_{ba}^{(1)}\theta_{ab}^{(1)}-2t^2\theta_{ab}^{(1)},\quad a\neq b\\
		\theta_{aa,aa,aa}^{(3)}={\theta_{aa}^{(1)}}^3-3t^2\theta_{aa}^{(1)}.
		\end{cases},\\
&&\quad ~=\theta_{ij}^{(1)} \theta_{kl}^{(1)} \theta^{(1)}_{mn}-\(\text{all the possible contractions}\)\equiv [\theta_{ij}^{(1)} \theta_{kl}^{(1)} \theta^{(1)}_{mn}].
\eea 
So the highest level sector can also be understood as the observable in the ``normal order". 
Applying this rule of decomposition to the single-trace observables we get
\bea 
&&\Tr H=\Tr \theta^{(1)},\\
&&\Tr H^2=\langle \Tr H^2\rangle+\sum_{ij}\theta_{ij}^{(2)}=\langle \Tr H^2\rangle+\(\Tr {\theta^{(1)}}^2-N^2 t^2\),\\
&&\Tr H^3=3\langle \Tr H^2\rangle \Tr\theta^{(1)}+\(\Tr({\theta^{(1)}}^3)-3Nt^2\Tr \theta^{(1)}\),\\
&&\Tr H^4=\langle \Tr H^4\rangle+\(4N^2t^2\Tr(\theta^{(2)})+2t^2[\Tr \theta^{(1)}\Tr \theta^{(1)}]\)+[\Tr  {\theta^{(1)}}^4],\\
&&\dots \nonumber
\eea 
where the normal ordered terms are explicitly given by
\bea 
&&[\Tr \theta^{(1)}\Tr \theta^{(1)}]=\Tr \theta^{(1)}\Tr \theta^{(1)}-Nt^2,\\
&&[\Tr {\theta^{(1)}}^4]=\Tr {\theta^{(1)}}^4-\( \langle \Tr H^4\rangle+\(4N^2t^2\Tr(\theta^{(2)})+2t^2[\Tr \theta^{(1)}\Tr \theta^{(1)}]\)\).
\eea 
Like the Wick transformation in quantum field theory,  the normal order or the highest level sector can be defined as
\bea 
[f(H)]=e^{-\frac{t^2}{2}\Tr\(\frac{\delta}{\delta H_{ij}}\frac{\delta}{\delta H_{ji}}\)}f(H),
\eea 
or we can introduce the formal integral 
\bea \label{gnormal}
[f(H)]=\int  \d \tilde{H}\,e^{\frac{1}{2t^2}\Tr\tilde{H}^2} e^{ -\frac{1}{t^2}\Tr(\theta^{(1)}\tilde{H})}e^{\frac{1}{2t^2}\Tr {\theta^{(1)}}^2} f( \tilde{H})-f(0),
\eea 
which is more convenient sometimes. Therefore we can rewrite the decomposition as
\bea 
f(H)=[f(H)]+\sum_k Con_k,
\eea 
where the $Con_k$ means choosing all possible $k$ pairs of matrix elements from $[f(H)]$ and replacing each pair $H_{ab}H_{cd}$ with its expectation value $\langle H_{ab}H_{cd}\rangle$. It implies the identification
\bea 
[f(H)]=\Theta_p,\quad Con_k=\Theta_{p-2k}.
\eea 
For these single-trace observables, in the large $N$ limit their correlation functions factorize so the dominant sector is always the self-averaged sector. The more interesting observable is $z(\im T)$ whose expectation value is
\bea 
\lim_{N\rightarrow \infty } \langle z (\im T)\rangle &=& \sum_{k=0} \frac{(\im T)^{2k}}{(2k)!}\langle \Tr H^{2k}\rangle=N\sum_{k=0} \frac{(\im T \sqrt{N t^2})^{2k}}{(2k)!}C_k\\
&=& N\frac{J_1(2 \alpha T )}{\alpha T }\sim 0,\quad \text{when }T>>1,
\eea 
where $C_k$ is famous Catalan number and $\alpha=\sqrt{N t^2}$.
So in the late time, the non-self-averaged sector becomes important. The lowest sector can be simply obtained  by expanding $z$ and picking the term with $\theta^{(1)}$\footnote{There is a $1/N$ in front because one of the summation of indexes gives the trace of $\theta^{(1)}$ instead of a factor of N. For example $\sum_{a,i,j,k,m}\theta_{ai}^{(1)}\langle H_{ij}H_{jk}H_{km}H_{ma}\rangle=\Tr\theta^{(1)}$.}:
\bea
\Theta_1&=& \Tr\theta^{(1)}\frac{1}{N}\(N \im T+\frac{(\im T)^3}{3!}3\times \langle \Tr H^2\rangle+\frac{(\im T)^5}{5!}5\times \langle \Tr H^4\rangle+\dots\)\\
&=&\im {\Tr\theta^{(1)}} \frac{J_1(2\alpha T)}{\alpha}\,.
\eea 

\begin{figure}[h]
\centering
  \includegraphics[scale=0.3]{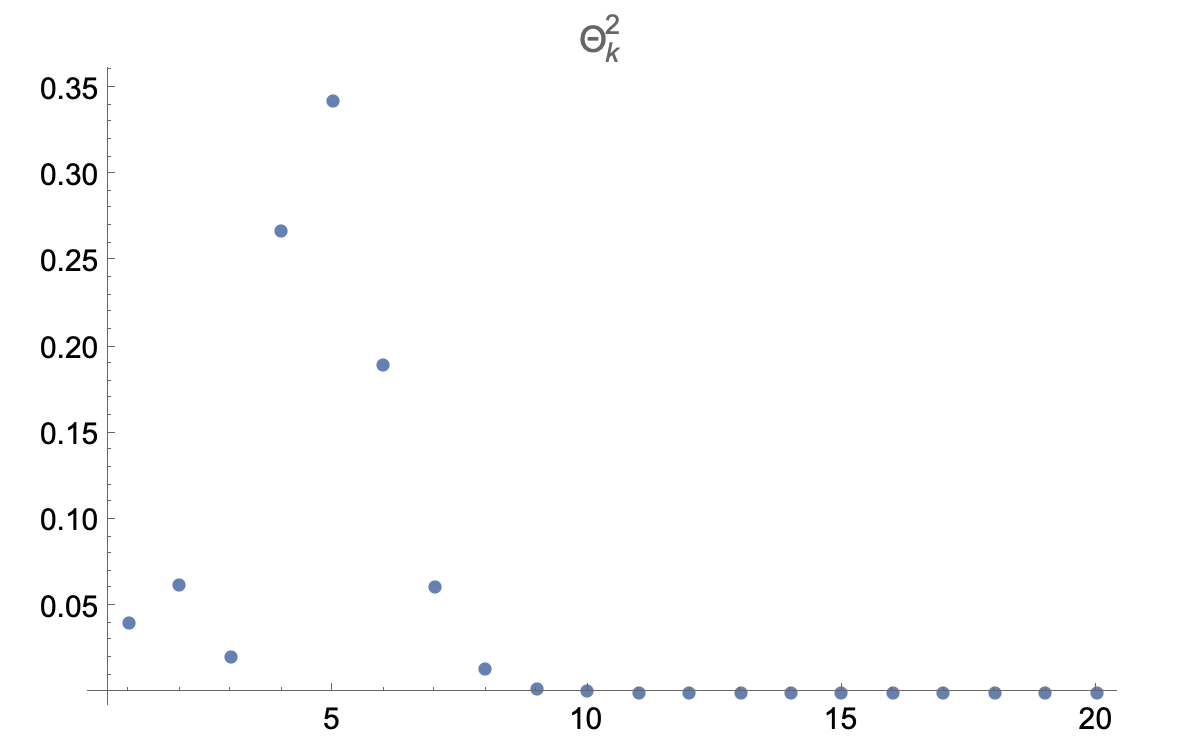}
  \caption{$\alpha=1,T=3.$}\label{RMT3}
\end{figure}

\begin{figure}[h]
\begin{subfigure}{.5\textwidth}
  \includegraphics[scale=0.32]{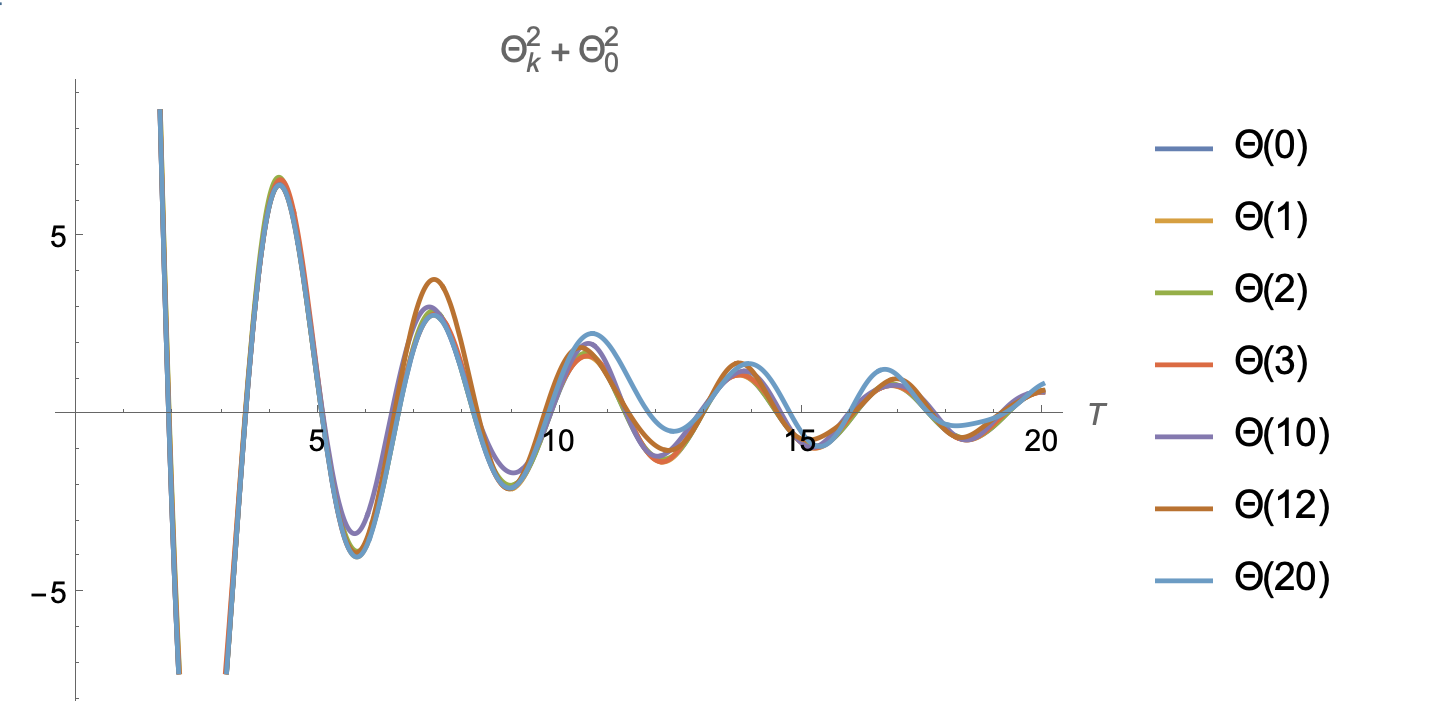}
  \caption{$N=100,\quad \alpha=1$}\label{RMT1}
 \end{subfigure}
\begin{subfigure}{.5\textwidth}
  \includegraphics[scale=0.32]{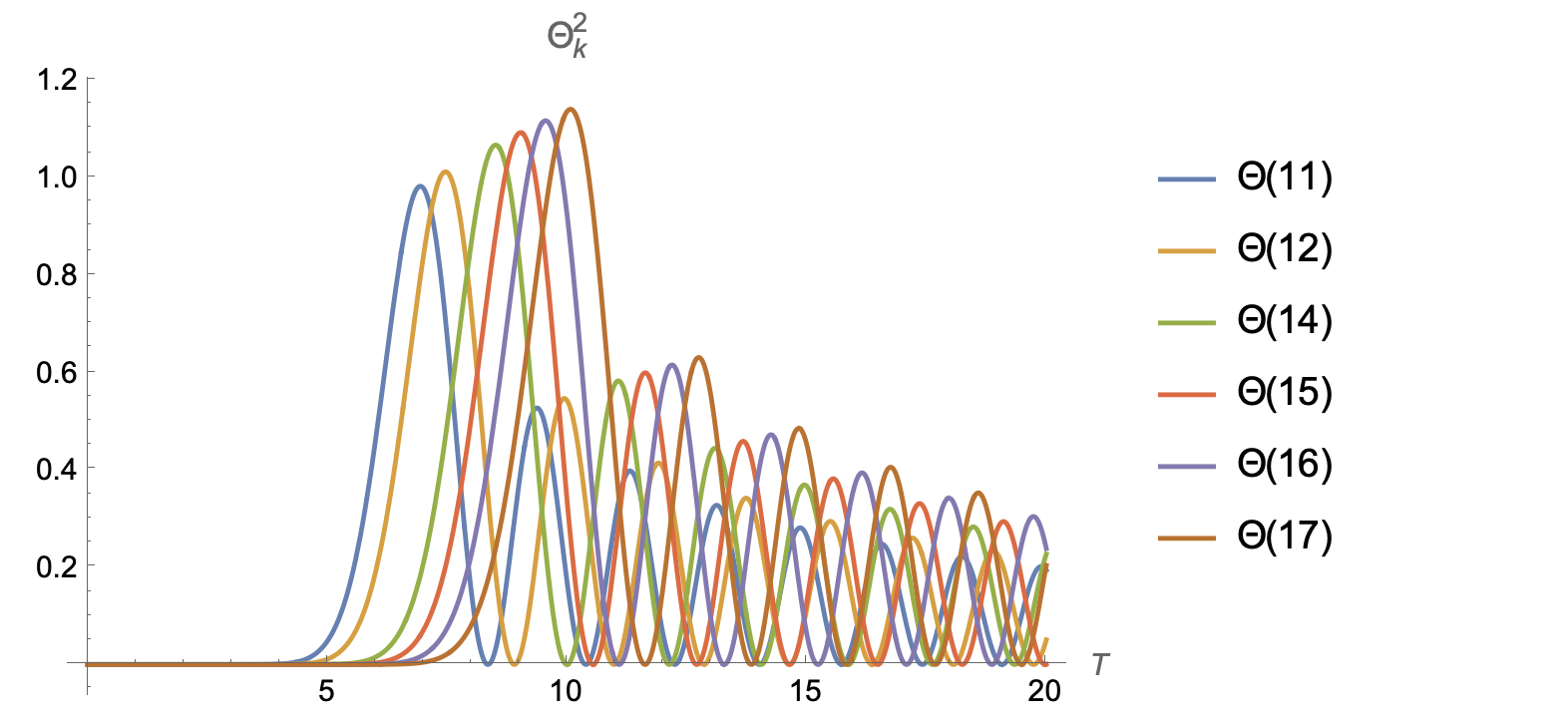}
  \caption{$N=100,\quad \alpha=1$}\label{RMT2}
  \end{subfigure}
  \caption{}
\end{figure}

Similarly we find that the next sector is \footnote{The factor $6\times 2$ comes from the adjacent terms like $\theta^{(2)}_{ab,ij}\langle H_{ij}H_{ik}H_{km}H_{mn}\rangle $ and factor $3$ comes from the pairs like $\theta^{(2)}_{ab,cd}\langle H_{bi}H_{ic}H_{dk}H_{ka}\rangle$.}
\bea 
\Theta_2 &=&\Tr \theta^{(2)}\frac{1}{\alpha^2}\(\frac{(\im T \alpha )^2}{2}+\frac{(\im T \alpha)^4}{4!}\times 4+\frac{(\im T \alpha)^6}{6!}\times (6*2+3)+\dots\),\\
&=&-\Tr\theta^{(2)}\frac{J_2(2T \alpha)}{\alpha^2},
\eea 
where we have dropped the terms  $\Tr\theta^{(1)}\Tr\theta^{(1)}$ because they are suppressed by $1/N$. Comparing with the known results\footnote{for example see \cite{Blommaert:2021fob}} of the wormhole contribution to $\langle z(\im T_1) z(\im T_2)\rangle_c$
\bea 
\langle z(\im T_1) z(\im T_2)\rangle_c=\sum_{l=0}^{\infty}(l+1)(-1)^{l+1}J_{l+1}(2\alpha T_1)J_{l+1}(2\alpha T_2),
\eea 
we will show in the Appendix \ref{matrix} that
\bea \label{tk}
\Theta_k=(\im)^k \frac{J_{k}(2\alpha T)}{\alpha^{k}}\Tr\theta^{(k)},\quad  k>0.
\eea 
We plot $\langle \Theta_0^2\rangle+\langle \Theta_k^2\rangle$  in Fig.\ref{RMT1} and $\langle \Theta_k^2\rangle$  in Fig.\ref{RMT2}. The result is very interesting. We see that every curve has the typical slop, ramp and plateau regimes. Another interesting fact is that only the first few sectors  contribute to the slop and ramp regions.
For example, adding the first 20 sectors we find that the ramp region is roughly located at $[2.5/\alpha,4/\alpha ]$ and we plot the contribution of each sector in Fig. \ref{RMT3}. Actually including the first 10 sectors is a very good approximation
\bea 
\frac{\sum_{i=1}^{10}\langle \Theta_i^2\rangle}{\sum_{i=1}^{20}\langle \Theta_i^2\rangle}=0.999974.
\eea 

 Therefore if we only focus on the ramp which is supposed to relate to wormholes we only need to include the first 10 non-self-averaged sectors. In this sense we may call $\Theta=\sum_{i=1}^{10}\Theta_{i}$ the half-wormhole of $z$. This is similar to the half-wormhole of the simple exponential observable \eqref{expy} in the regime $\beta^2 t^2<1$. We can follow the same procedure to study the decomposition and the half-wormhole of $z^2$. But it is very cumbersome and we expect the its behavior is similar to the exponential observable.

\section{Beyond Gaussian distribution or the generalized CGS model}
\begin{figure}[h]
\centering
  \includegraphics[scale=0.5]{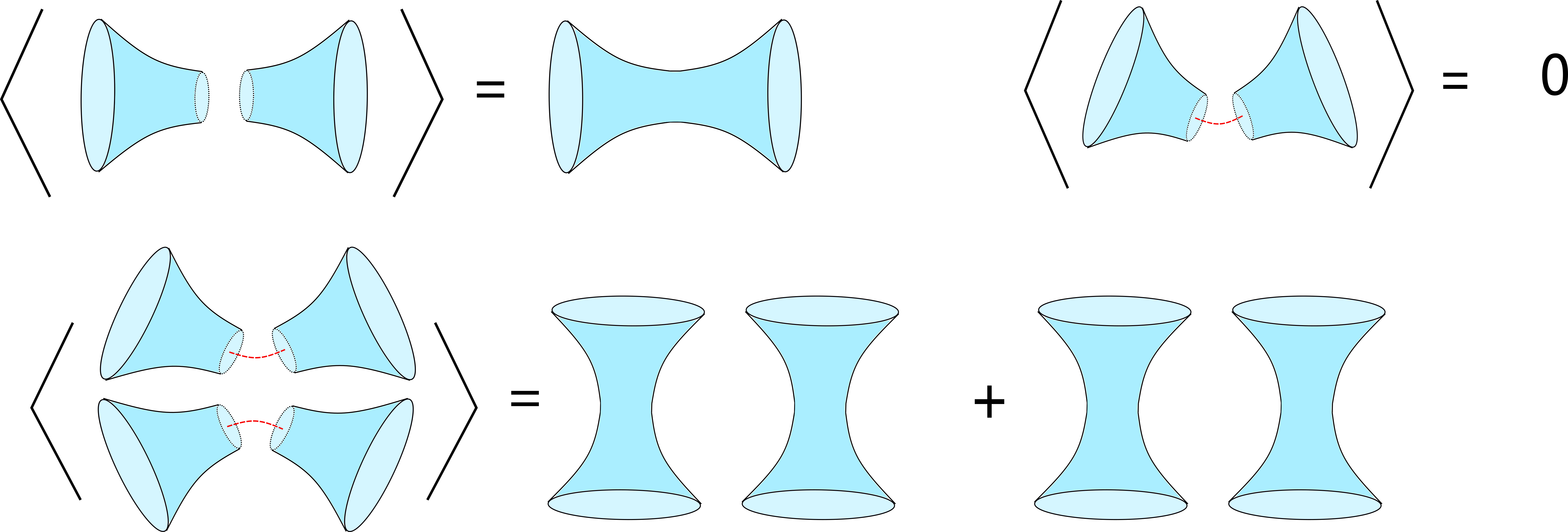}
  \caption{Illustrations of the metric \eqref{metric}}\label{muhalf1}
  \includegraphics[scale=0.5]{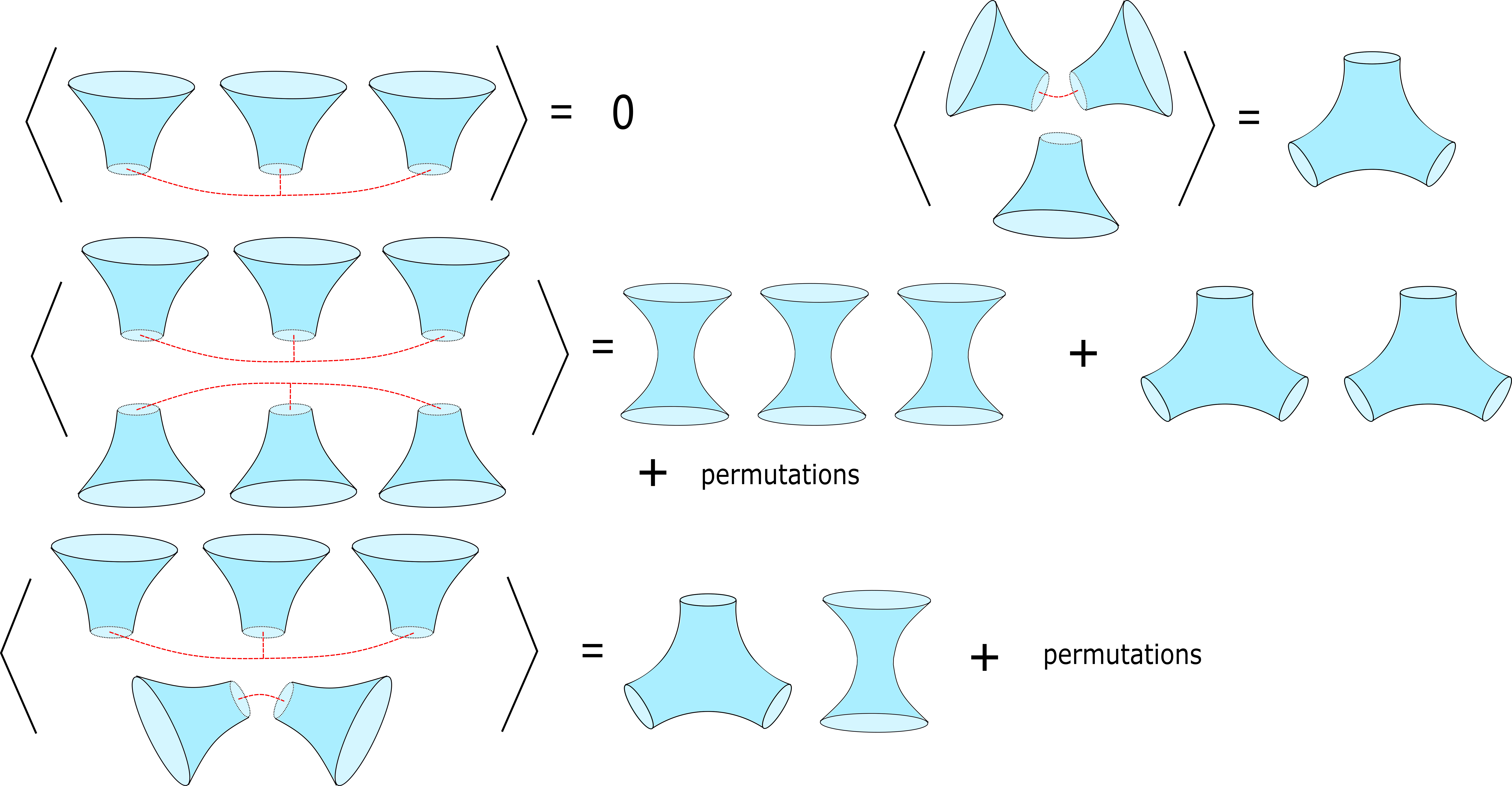}
  \caption{Illustrations of the metric \eqref{metric}}\label{muhalf2}
\end{figure}

One of simplest way to go beyond CGS model is again starting from the MM model but including connected spacetimes with other topologies in the Euclidean path integral. So the next simplest case beyond CGS model is the Disk-Cylinder-Pants model. Let the amplitudes of the disk, cylinder and pants to be
\bea 
\langle \hat{Z}\rangle=\kappa_1,\quad \langle \hat{Z}^2\rangle-\langle \hat{Z}\rangle^2=\kappa_2,\quad \langle \hat{Z}^3\rangle-3\langle \hat{Z}\rangle\langle \hat{Z}^2\rangle+2\langle \hat{Z}\rangle^2=\kappa_3.
\eea 
The generating function is
\bea 
\langle e^{u \hat{Z}}\rangle=\exp\(u\kappa_1+\frac{u^2 \kappa_2}{2!}+\frac{u^3 \kappa_3}{3!}\),
\eea 
so we can also identify $\hat{Z}$ as a random variable albeit with a very complicated PDF. We can simply think of the distribution is defined by the same generating function. In \cite{Peng:2022pfa} we introduce the connected correlators to decompose $\langle Z^n e^{\im k Z}\rangle$ for example
\bea 
&&\langle Z e^{\im k Z}\rangle=\langle Z\rangle\langle e^{\im k Z}\rangle+\langle Z e^{\im kZ}\rangle_{\text{c}}\,,\label{x2d1}\\
&&\langle Z^2 e^{\im k Z}\rangle=\langle Z^2\rangle \langle e^{\im k Z}\rangle+2\langle Z\rangle\langle Ze^{\im k Z}\rangle_c+\langle Z^2e^{\im k Z}\rangle_{\text{c}},\label{x2d}\\
&&\langle Z^3 e^{\im k Z}\rangle=\langle Z^3\rangle \langle e^{\im k Z}\rangle+3\langle Z^2\rangle\langle Ze^{\im k Z}\rangle_c+3\langle Z\rangle\langle Z^2e^{\im k Z}\rangle_c+\langle Z^3e^{\im k Z}\rangle_{\text{c}}\,,\label{x2d3}\\
&&\dots \nonumber
\eea 
such that using the trick \eqref{trick} we can decompose $Z_\alpha^n$ into different sectors which are exactly like \eqref{de1}-\eqref{de4}. In other words, the number basis or $\{\theta^{(i)}\}$ is still the basis for decomposition. But $\{\theta^{(i)}\}$ should be determined from the recursion relations \eqref{de1}-\eqref{de4}. For example, in the Disk-Cylinder-Pants model the first few $\theta^{(i)}$ are
\bea 
\theta^{(1)}=Z-\kappa_1,\quad \theta^{(2)}={\theta^{(1)}}^2-\kappa_2,\quad \theta^{(3)}={\theta^{(1)}}^3-3\kappa_2\theta^{(1)}-\kappa_3.
\eea 
Because of the inclusion of new wormholes, the pants, the basis is not orthogonal anymore in the sense
\bea 
\langle \theta^{(i)}\theta^{(j)}\rangle \neq \delta_{ij}.
\eea 
It is easy to find that
\bea 
\langle \theta^{(1)}\theta^{(2)}\rangle=\kappa_3,\quad \langle \theta^{(2)}\theta^{(3)}\rangle=6 \kappa_2 \kappa_3.
\eea 
Moreover the matrix $M^{(3)}_{ij}=\langle \theta^{(i)}\theta^{(j)}\rangle,i,j=1,2,3$ is 
\bea \label{metric}
M^{(3)}=\left(
\begin{array}{ccc}
 \kappa_2 & \kappa_3 & 0 \\
 \kappa_3 & 2 \kappa_2^2 & 6 \kappa_2 \kappa_3 \\
 0 & 6 \kappa_2 \kappa_3 & 6 \kappa_2^3+9 \kappa_3^2 \\
\end{array}
\right).
\eea 
Naturally $\theta^{(i)}$ can be understood as the $i$-linked half wormhole as shown in Fig.\ref{muhalf1} and \ref{muhalf2}\par 
\subsection{Toy statistical model}
We start from the simplest operator
\bea \label{y1}
Y=\sum_i \left( X_i-\langle X_i\rangle\right).
\eea 
The modification starts to show up in 
\bea 
Y^3=\Delta_0+\Delta_1+\Delta_3,
\eea 
where
\begin{align}
	&\Delta_{0}=\langle Y^{3}\rangle,\\
	&\Delta_{1}=3N\kappa_{2}\sum\theta^{(1)}_{i},\\
	&\Delta_{3}=\sum_{i}\theta^{(3)}_{i}+3\sum_{i\neq j}\theta^{(2)}_{i}\theta^{(1)}_{j}+\sum_{i\neq j\neq k}\theta^{(1)}_{i}\theta^{(1)}_{j}\theta^{(1)}_{k}.
\end{align}
In this special case since $\langle \theta^{(1)}\theta^{(3)}\rangle=0$, there is no cross terms in $\langle Y^6\rangle$  
\bea 
&&\langle Y^6\rangle=\sum_i \langle \Delta_i^2\rangle,\\
&&\langle \Delta_0^2\rangle\sim N^2\kappa_3^2,\quad \langle \Delta_1^2\rangle\sim N^3\kappa_2^3,\quad \langle \Delta_3^2\rangle\sim N^3\kappa_2^3+N^2\kappa_3^2,
\eea 
where we only keep the possible leading terms. When $\k_2,\k_3\sim \mathcal{O}(1)$, the operator $Y^3$ is not self-averaged and the effect of $\kappa_3$ is negligible. The interesting case is when $N^2\kappa_3^2>>N^3 \kappa_2^3$ so that we have the approximation
\bea 
Y^3\approx \langle Y^3\rangle+\Delta_3,
\eea 
which is the analog of \eqref{z2h}.

\subsection{0-SYK model}
Let us reconsider the 0-SYK model but assume the random couplings satisfying 
\bea 
&&\langle J_{i_1\dots i_q}\rangle=0,\quad \langle J_{i_1\dots i_q}J_{j_1\dots j_q}\rangle=\kappa_2\delta_{i_1j_1}\dots \delta_{i_qj_q},\quad \kappa_2=\tau^2\frac{(q-1)!}{N^{q-1}}\ ,\\
&&\langle J_AJ_BJ_C\rangle=\kappa_3 \delta_{ABC},
\eea 
we will determine the scaling of $\kappa_3$ in a moment.
Then the averaged quantity is
\bea 
\langle z^3\rangle=\int \d^{3N}\psi \,  e^{ \kappa_2\sum_{A}(\psi_A^1\psi_A^2+ \psi_A^1\psi_A^3+\psi_A^2\psi_A^3)}e^{\im^{3q}\k_3 \sum_A \psi_A^1\psi_A^2\psi_A^3},
\eea 
which can be computed by introducing the collective variables 
\bea 
G_{ab}&=&\frac{1}{N}\sum_{i}\psi_i^a\psi_i^b,\quad (a,b)=(1,2),(1,3),(2,3),\\
G_3&=&\frac{1}{N}\sum_{i<j}\psi_i^{1}\psi_j^{1}\psi_i^{2}\psi_j^{2}\psi_i^{3}\psi_j^{3},\\
\sum_A\psi_A^a\psi_A^b&=&\frac{N^q}{q!}G_{ab}^q,\quad \sum_A \psi_A^1\psi_A^2\psi_A^3=\frac{N^{q/2}}{(q/2)!}G_3^{q/2},
\eea 
to rewrite $\langle z^3\rangle$ as
\bea 
\langle z^3\rangle &=& \int [\frac{\d G_{ab}\d\S_{ab}}{2\pi\im /N}]\int \frac{\d G_3 \d \S_3}{2\pi\im /N}\, e^{\frac{N}{q}\(\tau^2 \sum G_{ab}^q+2\gamma^3G_3^{q/2}\)}e^{-N(\sum\S_{ab}G_{ab}+\S_3G_3)}\nonumber \\
&&\quad \int \d^{3N}\psi e^{\Sigma_{ab}\sum_i \psi_i^a\psi_i^b+\S_3 \sum_{i<j}\psi_i^{1}\psi_j^{1}\psi_i^{2}\psi_j^{2}\psi_i^{3}\psi_j^{3}}\\
&=& \int [\frac{\d G_{ab}\d\S_{ab}}{2\pi\im /N}]\int \frac{\d G_3 \d \S_3}{2\pi\im /N}\, e^{\frac{N}{q}\(\tau^2 \sum G_{ab}^q+2\gamma^3G_3^{q/2}\)}e^{-N(\sum\S_{ab}G_{ab}+\S_3G_3)}\S_3^{N/2}\nonumber \\
&=&\int \frac{\d G_3 \d \S_3}{2\pi\im /N}\, e^{\frac{N}{q/2}\gamma^3G_3^{q/2}}e^{-N\S_3G_3}\S_3^{N/2}=\gamma^{3p}m_p,
\eea 
where $m_p$ is defined in \eqref{mp} and
\bea 
\gamma^3\equiv\im^{3q}\k_3\frac{N^{q/2-1}}{(q/2-1)!},\quad \gamma\sim \mathcal{O}(1),
\eea 
thus
\bea 
\kappa_3 \sim \frac{(q/2-1)!}{N^{q/2}-1}.
\eea 
Recall that
\bea 
z^3=\sum'_{A,B,C}\text{sgn}(A)\text{sgn}(B)\text{sgn}(C)J_{A_1}J_{B_1}J_{C_1}\dots J_{A_p}J_{B_p}J_{C_p}.
\eea 

In general decomposing $z^3$ is still very complicated. Let us consider some simple examples. If $p=2$, then
we have 
\bea 
	z^{3}&=&\sum'_{A}\text{sgn}(A)J_{A_{1}}^{3}J_{A_{2}}^{3}+3\sum'_{A,B,A_i\neq B_i}\text{sgn}(B)J_{A_{1}}^{2}J_{A_{2}}^{2}J_{B_{1}}J_{B_{2}}\nn
	&&+\sum'_{A,B,C,A_i\neq B_i\neq C_i}\text{sgn}(A)\text{sgn}(B)\text{sgn}(C)J_{A_{1}}J_{A_{2}}J_{B_{1}}J_{B_{2}}J_{C_{1}}J_{C_{2}},
	\eea
and there are seven different sectors. A simple way to derive the explicit expression of each sector is to first decompose each $J_A^n$ as \eqref{de1}-\eqref{de4}:
\bea 
J_A=\theta_A^{(1)},\quad J_A^2=\theta_A^{(2)}+(\kappa_2)_A,\quad J_A^3=\theta_A^{(3)}+(\kappa_3)_A +3(\kappa_2)_A \theta_A^{(1)},
\eea
then collect the terms in the same sector:
\bea 
&& \Delta_0=\kappa_3^2\sum' \text{sgn}(A)=m_p\kappa_3^2=\langle z^3\rangle,\\
&& \Delta_1=3\kappa_3\sum'_A\text{sgn}(A)\(\theta_{A_1}^{(1)}(\kappa_2)_{A_2}+(\kappa_2)_{A_1}\theta_{A_2}^{(1)}\),\\
&&\Delta_2=(6+3M_p)\kappa_2^2\sum'_A\text{sgn}(A)\theta_{A_1}^{(1)}\theta_{A_2}^{(1)},\\
&&\Delta_3=\sum'_A\text{sgn}(A)\(\theta_{A_1}^{(3)}(\kappa_3)_{A_2}+(\kappa_3)_{A_1}\theta_{A_2}^{(3)}\),\\
&&\Delta_4=3\kappa_2\sum'_A\text{sgn}(A)\(\theta_{A_1}^{(3)}\theta_{A_2}^{(1)}+\theta_{A_1}^{(1)}\theta_{A_2}^{(3)}+\sum_{B,B\neq A_1,A_2} \theta_B^{(2)}\theta_{A_1}^{(1)}\theta_{A_2}^{(1)}\),\\
&&\Delta_5=0,\quad\Delta_6=\sum'_{A}\text{sgn}(A)\theta_{A_{1}}^{(3)}\theta_{A_{2}}^{(3)}+3\sum'_{A,B,A_i\neq B_i}\text{sgn}(B)\theta_{A_{1}}^{(2)}\theta_{A_{2}}^{(2)}\theta_{B_{1}}^{(1)}\theta_{B_{2}}^{(1)}\nn
	&&+\sum'_{A,B,C,A_i\neq B_i\neq C_i}\text{sgn}(A)\text{sgn}(B)\text{sgn}(C)\theta_{A_{1}}^{(1)}\theta_{A_{2}}^{(1)}\theta_{B_{1}}^{(1)}\theta_{B_{2}}^{(1)}\theta_{C_{1}}^{(1)}\theta_{C_{2}}^{(1)},
\eea 
where $M_p=\frac{(pq)!}{p!(q!)^{p}}$. Now we are ready to compute $\langle \Delta_i\Delta_j\rangle$ using the relation \eqref{metric}. It turns out that different sectors are still orthogonal for this case:
\bea 
&&\langle\Delta_{0}^2\rangle=m_{p}^{2}\kappa_{3}^{4},\quad \langle \Delta_{1}^2\rangle=18M_p\k_3^2\k_2^3,\\
&&\langle \Delta_2^2\rangle=(6+3M_p)^2M_p\k_2^6,\quad \langle \Delta_3^2\rangle=2M_p\k_3^2(6\k_2^3+9\k_3^2),\\
&&\langle \Delta_4^2\rangle=18M_p\k_2^3(6\k_2^3+9\k_3^2)+18M_p(2M_p-2)\k_2^6,\\
&&\langle \Delta_6^2\rangle=M_p(6\k_2^3+9\k_3^2)^2+9M_p(M_p-1)4\k_2^6+9 (m_p^2-M_p)\k_3^4\\
&&\qquad +6M_p(M_p-1)(M_p-2)\k_2^6.
\eea 
In large $N$ limit the relevant parameters have the following asymptotic behaviors
\begin{align}
	\kappa_{2}\sim \frac{(N/2-1)!}{N^{N/2-1}},\quad \kappa_{3}\sim\frac{(N/4-1)!}{N^{N/4-1}},\quad \k_3^2\sim \kappa_2^3 m_p e^N,\quad M_p\sim m_p^2 \sqrt{N}\label{k3},
\end{align}
then the approximation can be given as 
\begin{align}
	z^{3}\sim \Delta_{3}+ \Delta_{6}.
\end{align}
In general we find that when $p<<N (\text{or } q>>1)$, $z_3$ is not self-averaged, i.e. the wormhole does not persists, but the (three-linked) half-wormhole emerges. This fact can be intuitively understood as the following. In this limit because of the scaling \eqref{k3}, the three-mouth-wormhole amplitude is favored thus the possible dominate sectors are $\Delta_0$, $\Delta_{3p-3} $ and $\Delta_{3p}$:
\bea 
&&\langle \Delta_0^2\rangle = m_p^2 \k_3^{2p},\quad
\langle \Delta_{3p}^2\rangle >\langle (\sum'_A\text{sgn}(A) \theta_{A_1}^{(3)}\dots \theta_{A_p}^{(3)} )^2\rangle\sim M_p \k_3^{2p},\\
&&\langle \Delta_{3p-3}^2\rangle >(\sum'_A\text{sgn}(A) \sum_i\theta_{A_1}^{(3)}\dots (\k_3)_{A_i}\dots \theta_{A_p}^{(3)} )^2\rangle\sim M_p \k_3^{2p},
\eea 
and since $M_p>>m_p^2$ we conclude that $z^3\approx \Delta_{3p-3}+\Delta_{3p}$. This is similar to the result obtained in section \ref{section0syk}. In the same limit, $z$ is not self-averaged neither while the half-wormhole emerges.

\section{Discussion}
In this paper we have generalized the factorization proposal introduced in \cite{Saad:2021uzi}. The main idea is to decompose the observables into the self-averaging sector and non-self-averaging sectors. We find that the contributions from different sectors have interesting statistics in the semi-classical limit. When the self-averaging sector survives in this limit the observable is self-averaging. An interesting phenomenon is the sector condensation meaning the surviving non-self-averaging trend to condense and in the extreme case only one non-self-averaging sector is left-over resembling the Bose-Einstein condensation. Then the half-wormhole saddle is naturally understood as the condensed sectors. We apply the this proposal to simple statistical model , 0-SYK model and random matrix model. Half-wormhole saddles are identified and they are in agreement with the known results. With our proposal we also show the equivalence between the results in \cite{Saad:2021uzi} and \cite{Saad:2021rcu}. We also studied multi-linked-half-wormholes and their relations. There are some future directions.
\subsection*{Sector condensation}
It is interesting to understand the sector condensation better. We expect that it is some criterion for an ensemble theory or a statistical observable to potentially have a bulk description. So it deserves to study it in other gravity/ensemble theories. Definitely the extreme case mimicking the Bose-Einstein condensation is the most interesting one. We have not understood when it will happen and could it be used as some order parameter.  We expect by studying the ``phase diagram" in the sector space we can obtain more information about the observables and systems.
\subsection*{Complex coupling and half-wormholes}
In \cite{Mukhametzhanov:2021hdi}, it shows that factorization is related to the complex couplings. In our approach, the complex coupling emerges as an auxiliary parameter to obtain the half-wormhole saddle. The trick here is similar to the one used by Coleman, Giddings and Strominger \cite{Coleman:1988cy,Giddings:1988wv,Giddings:1988cx}, where the non-local effect of spacetime wormhole is ``localized'' with a price of introducing random couplings. But the current analysis shows that this is only possible when ``Bose-Einstein" happens such that the dominant sector can be obtained from this trick. So it would be interesting to explore the relation between complex coupling and half-wormhole further using our approach. 
\subsection*{Relations to other factorization proposal}
Besides the half-wormhole proposal, there exists other proposals of factorization. For example, in \cite{Blommaert:2021fob} it shows two dimensional gravity can be factorized by including other branes in the gravitational path integral. These new branes corresponding to specific operators in the dual matrix model. From the point of view of our approach, inserting operators may be related to adding back the contributions from non-self-averaging sectors. 
 In \cite{Cheng:2022nra}, it is argued that factorization can be restored by adding other kinds of asymptotic boundaries corresponding to the degenerate vacua. It is clear that from our approach, this is equivalent to introducing new random variables. It would be interesting to see how this changes the statistic of contribution form different sectors.

\appendix
\begin{acknowledgments}
We thank Cheng Peng for valuable discussion and comments on a early version of the draft.  We thank many of the members of KITS for interesting related discussions. 
JT is supported by  the National Youth Fund No.12105289 and funds from the UCAS program of special research associate.
\end{acknowledgments}
\section{Details of \ref{matrixde}}
\label{matrix}
First let us rederive the non-self-averaged sectors of $z$ in a more systematic way. For simplicity let us set $t^2=1/N$. Defining
\bea
J_{ba}&=&\langle \frac{\delta z}{\delta H_{ab}}\rangle=\int \d H e^{-\frac{N}{2}\Tr H^2} \frac{\delta z}{\delta H_{ab}},\\
&=&N \int \d H e^{-\frac{N}{2}\Tr H^2} H_{ba} \, \Tr (e^{\im T H})\label{jab}=N\langle H_{ba}z\rangle
\eea 
then we can rewrite $\Theta_1$ as
\bea 
\Theta_1=\sum_{i,j}\theta_{ij}^{(1)}J_{ji}\equiv\langle \Tr(\theta^{(1)}\delta z)\rangle.
\eea 
By considering that in \eqref{jab} $H_{ba}$ has to contract with other $H_{ab}$ or by a argument of symmetry it is obvious 
\bea 
J_{ba}= \delta_{ab}J_{aa}
\eea 
thus
\bea 
J_{ba}=\delta_{ab} N\langle H_{aa}z\rangle=\delta_{ab}\langle \Tr H z\rangle, \quad \Theta_1=\sum_{i}\theta_{ii}^{(1)}J_{ii}=\Tr\theta^{(1)} \langle \Tr H z\rangle,
\eea 
where we have used the fact there is a permutation symmetry in the diagonal elements $\{H_{ii}\}$. Similarly the second non-self-averaged sector $\Theta_2$ can be written as
\bea 
\Theta_2=\frac{1}{2}\Tr\theta^{(2)}\langle (\Tr H^2-N)z\rangle.
\eea 
In general, it is
\bea \label{guessk}
\Theta_k=\frac{1}{k}\Tr\theta^{(k)}\langle [\Tr H^k]z\rangle,
\eea 
which simply means that $\{[\Tr H^k]\}$ is an orthogonal basis in the sense
\bea \label{noguess}
\langle [\Tr H^k][\Tr H^l]\rangle =k \delta_{kl}\,.
\eea 
Recall \eqref{gnormal} the generating function of the normal-ordered operator is 
\bea 
[G(u)]=\int  \d \tilde{H}\,e^{\frac{1}{2t^2}\Tr\tilde{H}^2} e^{ -\frac{1}{t^2}\Tr(H\tilde{H})}e^{\frac{1}{2t^2}\Tr {H}^2} \Tr e^{u \tilde{H}}.
\eea 
Therefore similar to the computation of \eqref{kk} we have
\bea 
&&\Big\langle [\Tr e^{u_L \tilde{H}}][\Tr e^{u_R \tilde{H}}]\Big\rangle=\int [\d H\d H_L \d H_R] e^{\frac{N}{2}(\Tr H^2_L+\Tr H^2_R+\Tr H^2)}e^{-N\Tr H(H_L+H_R)} \Tr e^{u_L H_L}\Tr e^{u_R H_R}\nonumber \\
&&=\int [\d H_L \d H_R]e^{-N\Tr H_L H_R}\Tr e^{u_L H_L}\Tr e^{u_R H_R}=\sum_k \frac{u_L^k}{k!}\frac{u_R^k}{k!} k, \label{kk2}
\eea 
where we have used the formal integral
\bea 
\int [\d H_L \d H_R]e^{-N\Tr H_L H_R}  [H_L]_{ij}[H_R]_{ji}=\frac{1}{N}\,. 
\eea 
By expanding both sides of \eqref{kk2} we get \eqref{noguess} as promised.
So the task is to compute the two-point correlation functions
\bea
\langle [\Tr H^n] z\rangle,\quad \text{or   }\quad\langle \Tr H^n z\rangle
\eea 
or more conveniently the generating function
\bea 
&&G(u)=\langle \Tr(e^{u H})z\rangle=\langle z(u)\rangle \langle z(\im T)\rangle+\sum_{l=0}^{\infty}(l+1)(-1)^{l+1}J_{l+1}(-2\im  u)J_{l+1}(2 T),\\
&&=N\frac{J_1(-2\im u)}{-\im u} \langle z(\im T)\rangle+\sum_{l=0}^{\infty}(l+1)(-1)^{l+1}J_{l+1}(-2\im  u)J_{l+1}(2 T).\label{gu}
\eea 
Expanding the generating function gives 
\bea 
&&\langle \Tr(H)z\rangle=\im J_{1}(2T),\quad \langle \Tr(H^2)z\rangle=N\langle z\rangle-2J_{2}(2T),\\
&&\langle \Tr(H^3)z\rangle=-3\im J_3(2T)+3\im J_1(2T),\\
&&\langle \Tr(H^4)z\rangle=2N\langle z\rangle-8 J_2(2T)+4 J_4(2T),\dots
\eea
which indeed lead to \eqref{tk}. \par 
It would be desired to derive a generating function of the normal ordered operators $[\Tr H^n]$ which has the integral form
\bea \label{gun}
[G(u)]=\int  \d \tilde{H}\,e^{\frac{1}{2t^2}\Tr\tilde{H}^2} e^{ -\frac{1}{t^2}\Tr(H\tilde{H})}e^{\frac{1}{2t^2}\Tr {H}^2} \Tr e^{u \tilde{H}}.
\eea 
Note that \eqref{gun} describes a GUE model coupled with an external source. As shown in \cite{Blommaert:2021fob} it can be rewritten as
\bea 
[G(u)]&=&\int \prod_i \d \tilde{\lambda}_i e^{\frac{1}{2t^2}\sum \tilde{\lambda}_i-\frac{1}{t^2}\sum \tilde{\lambda}_i\lambda_i+\frac{1}{2t^2}\sum \lambda_i^2 }\frac{\Delta(\tilde{\lambda})}{\Delta(\lambda)}\sum_k e^{u \tilde{\lambda}_k}\\
&=&\sum_j e^{\frac{1}{2t^2}\sum {\lambda}_i^2-\frac{1}{2t^2}\sum_i(\lambda_i-\delta_{ji}t^2 u)^2}\prod_{i\neq j}^N(1+\frac{-ut^2}{\lambda_i-\lambda_j})\\
&=&e^{-\frac{t^2 u^2}{2}}\frac{1}{-ut^2}\oint_{H}\frac{\d w}{2\pi \im} \prod_{i=1}^N(1+\frac{-ut^2}{w-\lambda_i})e^{wu}.\label{gun1}
\eea 
Notice that in the large $N$ limit $[\Tr H^n]$ is a linear combination of single trace operator so we should expand each $1/(w-\lambda_i)$ into Taylor series and only keep terms with $\sum_i \lambda_i^k$ 
\bea 
[G(u)]=e^{-\frac{u^2}{2N}}(-\frac{N}{u})\oint_{H}\frac{\d w}{2\pi \im}\((1+\frac{-u/N}{w})^N+\frac{-u}{N}(1+\frac{-u/N}{w})^{N-1}\sum_k \frac{\sum_i \lambda_i^k}{w^{k+1}}\)e^{uw}\nonumber ,
\eea 
where we have substituted $t^2=1/N$. Sending $N$ to infinity  gives
\bea 
e^{-\frac{u^2}{2N}}\sim 1,\quad (1+\frac{-u/N}{w})^N\sim (1+\frac{-u/N}{w})^{N-1}\sim e^{-u/w},
\eea 
thus we arrive at the final result
\bea 
[G(u)]&=&(-\frac{N}{u})\oint_{H}\frac{\d w}{2\pi \im} e^{u(w-\frac{1}{w})}+\sum_{k=1} \Tr H^k\oint_{H}\frac{\d w}{2\pi \im} e^{u(w-\frac{1}{w})}\frac{1}{w^{k+1}}.
\eea 
These contour integral can be evaluated exactly by using the expansion
\bea 
e^{u(w-\frac{1}{w})}=\sum_{i=-\infty}^\infty w^kJ_k(2u),
\eea 
which leads to
\bea\label{gunf} 
[G(u)]=\frac{N}{u}J_1(2u)+\sum_k\Tr H^k J_k(2u).
\eea 
By expanding with respect to $u$, indeed we get the correct normal-ordered operators
\bea 
[G(u)]&=&N+u \Tr H+\frac{u^2}{2!}(\Tr H^2-N)+\frac{u^3}{3!}(\Tr H^3-3\Tr H)+\frac{u^4}{4!}\(\Tr H^4-4\Tr H^2+2N\)\nonumber \\
&+&\frac{u^5}{5!}\(\Tr H^5-5\Tr H^3+10\Tr H\)+\frac{1}{6!}\(\Tr H^6-6\Tr H^4+15\Tr H^2-5N\)\dots\,.
\eea 
We can also obtain a generating function of $\Theta_k$
\bea 
&&\langle [G(u)]z\rangle =\int \d H \int  \d \tilde{H}\,e^{\frac{N}{2}\Tr\tilde{H}^2} e^{ -N \Tr(H\tilde{H})} \Tr e^{u \tilde{H}}\Tr e^{\im T{H}}\\
&&=\frac{N}{u}J_1(2u)\langle z\rangle+\sum_k J_k(2u)\langle \Tr H^k \Tr e^{\im T H}\rangle 
\eea 
which unfortunately does not have a simple closed form but the ensemble average $\langle \Tr H^k \Tr e^{\im T H}\rangle$ can be computed with the generating function \eqref{gu}.

\end{document}